\documentclass{pasj00}
\begin{document}
\SetRunningHead{N. Ota et al.}{Chandra Observation of a Group of
  Galaxies HCG~80} 
\Received{2004/07/21}%{yyyy/mm/dd}
\Accepted{2004/08/22}%{yyyy/mm/dd}

\title{Chandra Observation of a Group of Galaxies HCG~80: \\
Does the Spiral-Only Group Have Hot Intragroup Gas?}

 \author{%
   Naomi \textsc{Ota},\altaffilmark{1}
   Umeyo \textsc{Morita},\altaffilmark{2}
   Tetsu \textsc{Kitayama},\altaffilmark{3} 
   and
   Takaya \textsc{Ohashi}\altaffilmark{2}}
 \altaffiltext{1}{Cosmic Radiation Laboratory, 
 RIKEN, 2-1 Hirosawa, Wako, Saitama 351-0198}
 \email{ota@crab.riken.jp}
 \altaffiltext{2}{Department of Physics, Tokyo Metropolitan University, 
   1-1 Minami-Osawa, Hachioji, Tokyo 192-0397}
 \altaffiltext{3}{Department of Physics, Toho University, 2-2-1 Miyama, 
   Funabashi, Chiba 274-8510}


\maketitle

\begin{abstract}
We present an analysis of Chandra X-ray observations of a compact
group of galaxies, HCG~80 ($z=0.03$). The system is a spiral-only group
composed of four late-type galaxies, and has a high-velocity dispersion
of 309 km s$^{-1}$.  With high-sensitivity Chandra observations, we
searched for diffuse X-ray emission from the intragroup medium
(IGM); however, no significant emission was detected.  We place a
severe upper limit on the luminosity of the diffuse gas as $L_{\rm X} <
6\times10^{40}$~erg s$^{-1}$.  On the other hand, significant emission
from three of the four members were detected. In particular, we
discovered huge halo emission from HCG~80a that extends on a scale of
$\sim 30$~kpc perpendicular to the galactic disk, whose X-ray
temperature and luminosity were measured to be $\sim 0.6$~keV and
$\sim 4\times10^{40}$~erg s$^{-1}$ in the 0.5--2 keV band, respectively.
It is most likely to be an outflow powered by intense starburst
activity. Based on the results, we discuss possible reasons for the
absence of diffuse X-ray emission in the HCG~80 group, suggesting that the
system is subject to galaxy interactions, and is possibly at an early
stage of IGM evolution.
\end{abstract}

\section{Introduction}

The majority of galaxies are found to reside in groups of galaxies
\citep{Tully_1987}, and the intragroup medium may dominate the total
baryon density of the local universe \citep{Fukugita_etal_1998}.
Groups of galaxies should thus provide useful probes of structure
formation in the universe, yet their physical nature is still highly
unclear [see \citet{Hickson_1997} and \citet{Mulchaey_2000} for
  reviews].  The detection of extended X-ray emission has at least
shown that a number of groups are gravitationally bound. The observed
X-ray luminosities and the inferred gas mass, however, exhibit
correlations with the gas temperature and galaxy velocity dispersion,
much steeper than those predicted from a simple self-similar model
\citep{Mulchaey_etal_1996, Ponman_etal_1996}. The X-ray emission also
tends to be suppressed in spiral-dominated groups
\citep{Osmond_Ponman_2004}, with a possible exception of HCG~57, from
which \citet{Fukazawa_etal_2002} detected extended thermal X-ray
emission with ASCA. Despite large observational errors, these facts
are likely to indicate a close link between galaxy evolution and the
properties of the intragroup medium.

Among the 109 galaxy groups observed with the ROSAT PSPC, extended
X-ray emission has been detected in at least 61 groups. No diffuse
emission was detected in 12 groups with only spiral members
\citep{Mulchaey_etal_2003}. Recently the GEMS project constructed a
large sample containing 60 groups based on the optical and ROSAT PSPC
catalogues \citep{Osmond_Ponman_2004}. The X-ray emission was detected
for three of the five spiral-only groups; however, due to the 
limited quality of the data, the origin of the emission was not
directly constrained, but was classified into hot halos of the
individual galaxies based on their threshold in the spatial extent of
$<60$ kpc.  They suggested, based on the anti-correlation between the
spiral fraction and the X-ray luminosity, that the presence of
detectable hot diffuse gas is strongly related to the galaxy
morphology. The possible absence of diffuse X-rays in the spiral-only
groups would thus imply that either the bound group should contain at
least one early-type galaxy, or the X-ray emission is preferentially
suppressed in the spiral-only groups for some reason.

To further explore the nature of diffuse emission from the groups of
galaxies, higher quality X-ray data are required.  In particular, a high
spatial resolution is crucial to separate the emission associated with
individual galaxies from the diffuse component.  Chandra and
XMM-Newton are the most suitable satellites for this purpose, yet
there have so far been few available observational results.
\citet{Belsole_etal_2003} recently reported on the detection of
diffuse X-ray emission from HCG~16 ($z=0.013$) with the XMM-Newton
EPIC cameras.  The temperature and the luminosity were measured to be
$0.49\pm0.17$ keV and $5.0\times10^{40}h_{70}^{-2}$ erg s$^{-1}$,
respectively. The result obeys the $L_{\rm X}-T$ relation obtained for
brighter galaxy groups, though it is located at the extreme faint end,
from which they suggested that HCG~16 is a bound system.

In this paper, we present a Chandra observation of a spiral-only group
of galaxies, HCG~80. This group has been one of the most plausible
candidates, among the known spiral-only groups, for the positive
detection of diffuse X-rays for the following reasons: (1) The high
line-of-sight velocity dispersion of $\sigma_{\rm v}=309$ km s$^{-1}$
implies a large gravitational potential; accordingly high X-ray
luminosity and temperature of $L_{\rm X}\sim
8.4\times10^{41}h_{70}^{-2}$~erg s$^{-1}$ and $kT\sim 0.8$~keV are
expected from the $\sigma_{\rm v}-L_{\rm X}$ and $\sigma_{\rm v}-T$
correlations \citep{Ponman_etal_1996}. The expected luminosity is
close to the $3\sigma$ upper limit from the previous ROSAT PSPC,
$L_{\rm X}=7.6 \times 10^{41}h_{70}^{-2}$ erg s$^{-1}$
\citep{Ponman_etal_1996}.  (2) A compact galaxy distribution (4
members within $1\arcmin.7$ diameter) has led to a judgment that the
HCG~80 galaxies are accordant members \citep{Arp_1997,
  Sulentic_1997}. (3) Two of the galaxies are classified as Im, which
may indicate the galaxy-galaxy interaction in the group.

In addition, previous observations suggested a causal link between the
starburst activity and the galaxy interactions [see
  \citet{Kennicutt_1998} for reviews].  Thus, a detailed spatial
analysis of individual galaxies with an arc-second resolution of
Chandra is a vital clue in constraining the connection between the
outflowing gas and the intragroup medium without suffering from the
contamination of point sources. The Chandra observations revealed
X-ray views of the extended halo emission from the nearby starburst
and normal galaxies in the dense environment. For example, the nature
of diffuse X-ray emission from NGC~253 and NGC~55, both of which
belong to the nearby spiral-only group, the Sculptor group, is considered 
separately in \citet{Strickland_etal_2002} and
\citet{Oshima_2003}. They exhibit observational evidence for galactic
outflow from the spirals powered by star-formation activity.
Strickland et al. (2004a, 2004b)
%\citet{Strickland_etal_2004a,Strickland_etal_2004b} 
studied ten
star-forming disk galaxies with the Chandra X-ray and H$\alpha$
imaging data. From the correlations between a variety of X-ray
quantities with multi-wavelength data, they quantitatively
investigated supernova feedback on the galactic scale.

Through such detailed studies on spiral galaxies, we expect to obtain a
new insight into the galaxy--IGM connection.  Previous ROSAT
observations showed that the groups of galaxies have a steeper
luminosity--temperature relation compared to that of clusters of galaxies,
which is noticeable for low temperature ($kT\lesssim 1$~keV) systems.
The preheating effect is suggested to be responsible for the
steepening (e.g., \cite{Ponman_etal_1996}); 
\citet{Helsdon_Ponman_2003} have proposed that spiral galaxies play a
comparable role to early types in gas heating.  However, it is still
ambiguous; investigations are needed to clarify it further.

This paper is organized as follows. In the next section, we describe
the Chandra observation of a galaxy group, HCG~80, and source detection
in the field. In section~\ref{sec:analysis_and_result}, we present
spatial and spectral analyses of HCG~80 member galaxies, and also
constrain the diffuse X-ray emission from the hot intragroup gas.  In
section~\ref{sec:discussion}, we discuss the properties of the member
galaxies and, in particular, the nature of the extended halo emission
discovered in the brightest member of the group. We then discuss
possible reasons for the absence of the strong X-ray emission from the
intragroup gas of HCG~80.

Throughout this paper we adopt $\Omega_{\rm M}=0.3$,
$\Omega_{\Lambda}=0.7$; $h_{70}\equiv H_0/(70~{\rm km\,
  s^{-1}\,Mpc^{-1}})=1$; $1\arcmin$ corresponds to 36.1~kpc at
$z=0.0299$.  The quoted errors indicate the 90\% confidence range,
unless otherwise stated. We use the solar abundance ratio of
\citet{Anders_Grevesse_1989}.

\section{Observation and Source Detection}

\subsection{Chandra Observation of HCG~80}

\begin{figure}
  \begin{center}
    \FigureFile(80mm,80mm){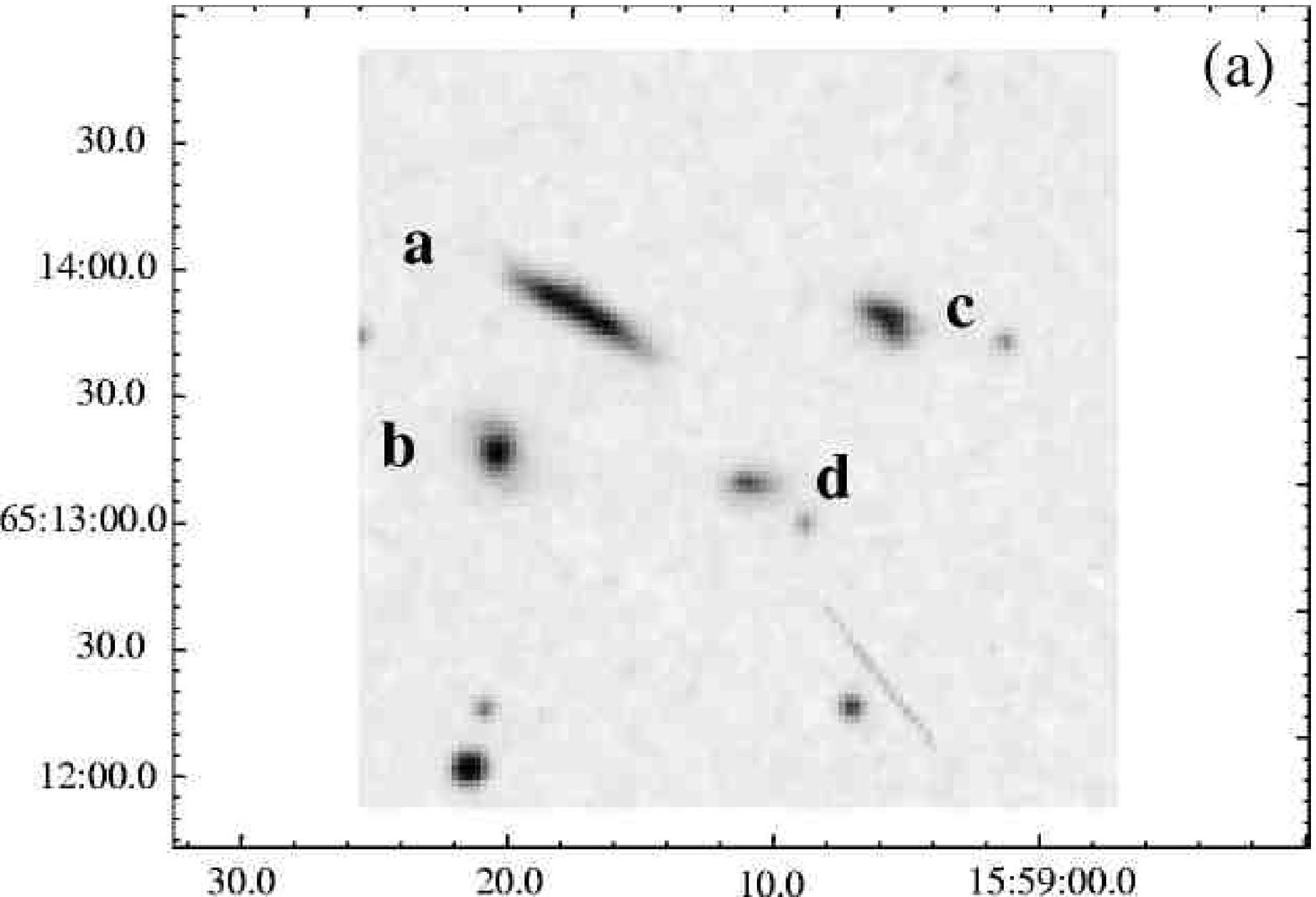}

    \FigureFile(80mm,80mm){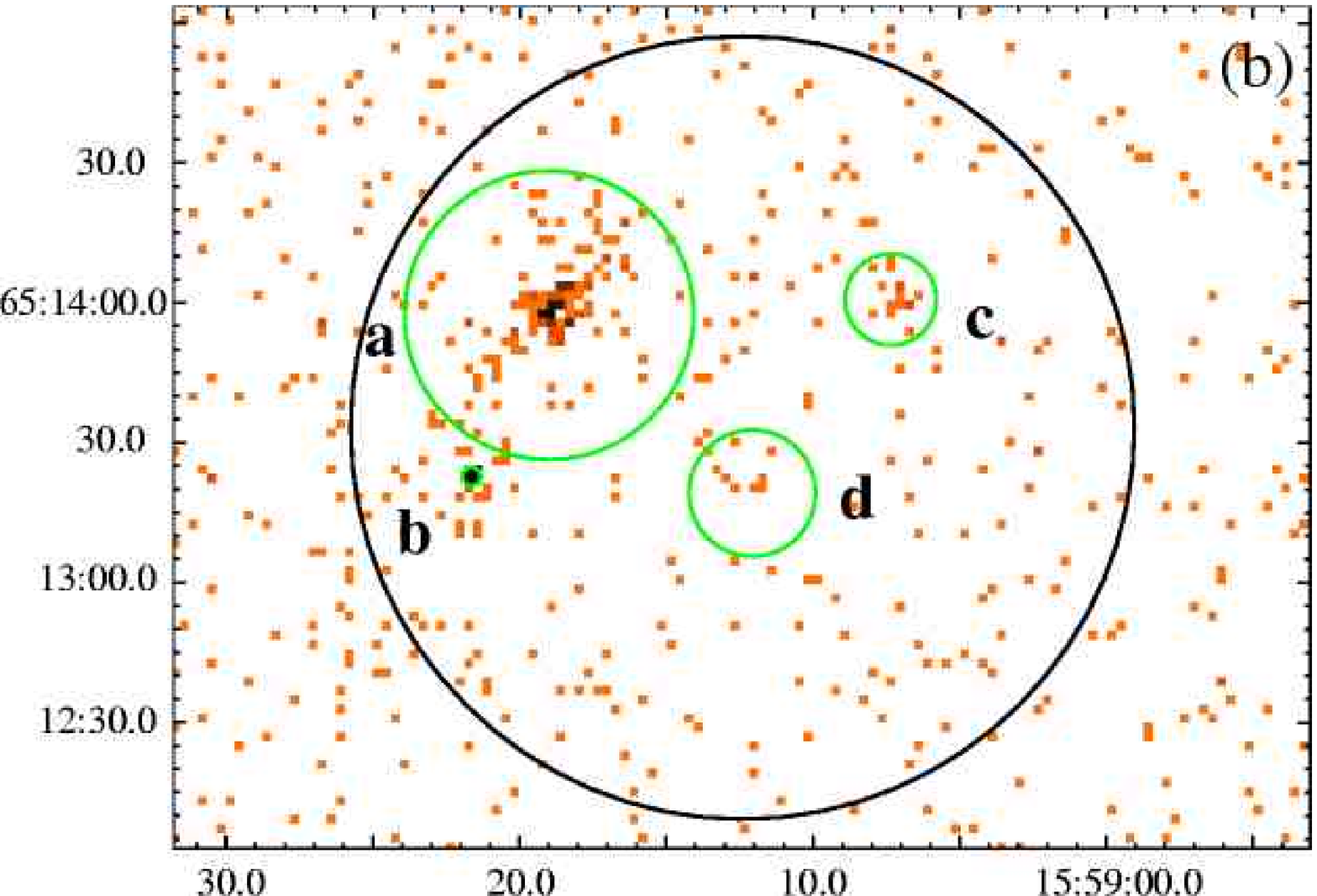}

    \FigureFile(80mm,80mm){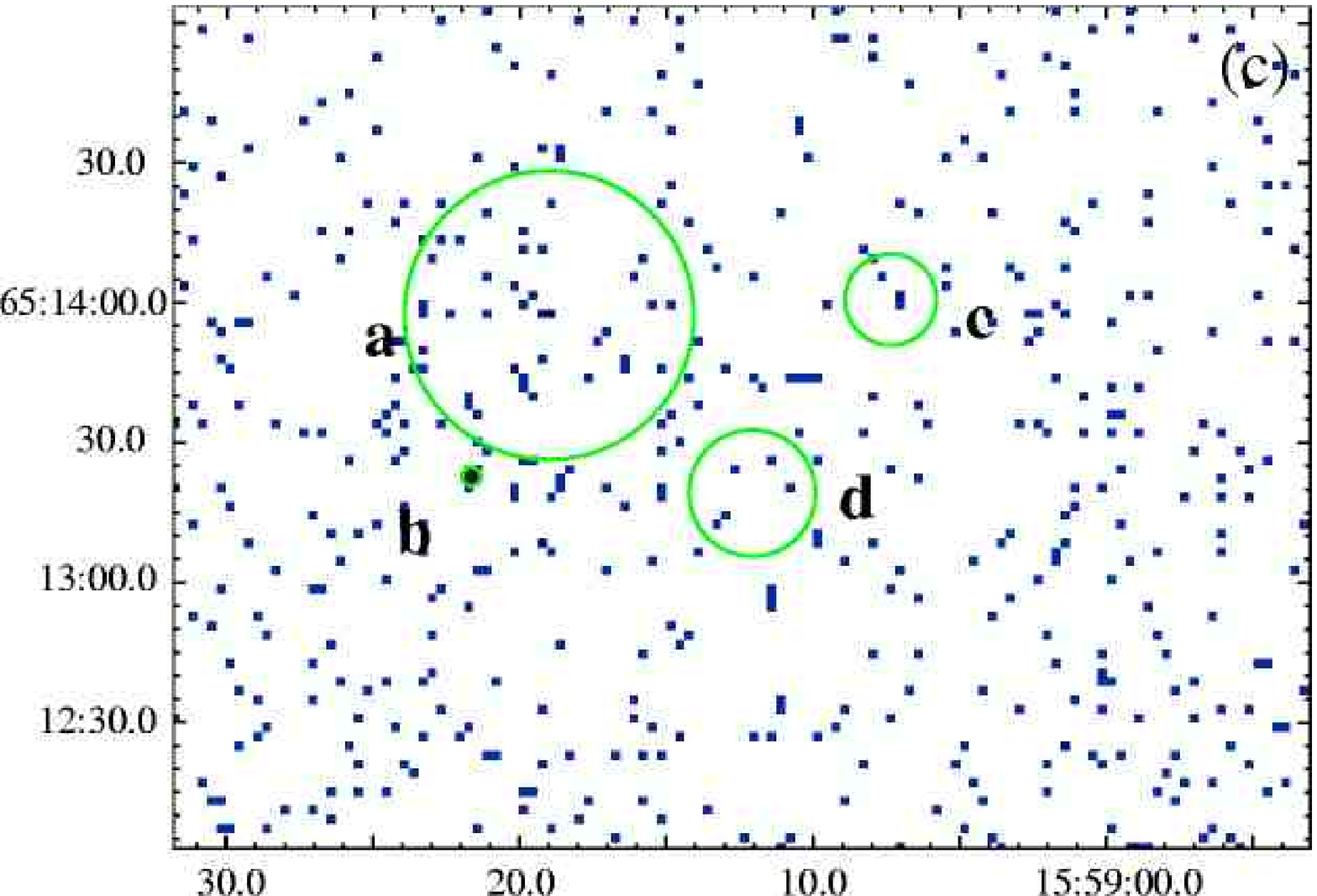}
  \end{center}
  \caption{Optical and X-ray images of the spiral-only group
 HCG~80. In the panel (a), the DSS image of the HCG~80 group is shown,
    where the four members are denoted by a--d. In panels (b)
    and (c), the Chandra X-ray images in the 0.5--2 keV and 2--7 keV
    bands are shown. The X-ray images are not smoothed, and are background
    inclusive. The image pixel size is $2\arcsec\times2\arcsec$.  
    The member galaxies, HCG~80a--d, are indicated by green
    circles of radii $r=31\arcsec$, $2\arcsec$, \timeform{9.8"}, and \timeform{13.6"},
    respectively.  In panel (b), the black color represents pixels 
    with an X-ray surface brightness higher than 
    10 photons per image pixel. The group region used to constrain
    the diffuse emission in subsection~\ref{subsec:diffuse_emission} is
    shown with a circle of radius $84\arcsec$.}\label{fig1}
\end{figure}

The group of galaxies HCG~80 consists of four late-type galaxies, HCG
80a--d, as shown in table~\ref{tab1} and figure~\ref{fig1}.  Their
optical properties are also summarized in table~\ref{tab1}.  We
observed HCG~80 with the Chandra Advanced CCD Imaging Spectrometer
(ACIS-S) detector on 2003 August 18 (PI: N. Ota). The pointing
coordinates were \timeform{15h59m12.30s}, \timeform{+65\circ13'33.0''}
(J2000), and the target was offset from the ACIS-S nominal aim point
with a Y-offset of $-1\arcmin$. The CCD temperature was $-120{\rm
  ^{\circ}C}$. The data reduction was performed using CIAO version
3.0.2 with CALDB version 2.25. We analyzed the light curve, and showed
that there is no period of a high background level exceeding $3\sigma$
above the mean quiescent background rates. Thus, the net exposure time
is 19712 s. We did not find any astrometry offset for the data.  Since
the data were taken with the VFAINT mode, the particle background was
reduced by screening out events with significant flux in border pixels
of the $5\times5$ event islands.

%%%%%%%%%%%%%%%%%%%%%%%%%%%%%%%%%%%%%%%%%%%%
%%%%%%%%%%%%%%%%%%%%%%%%%%%%%%%%%%%%%%%%%%%%

%%%%%%%%%%%%%%%%%%%%%%%%%%%%%%%%%%%%%%%%%%%%
\begin{table*}
\begin{center}
\caption{Optical properties of member galaxies.}
\label{tab1}
\begin{tabular}{lllllll}\hline\hline
Object & \multicolumn{2}{c}{Optical coords. (J2000)\footnotemark[*]} & $z$ & Diameters\footnotemark[$\dagger$] & \multicolumn{1}{c}{B\footnotemark[$\ddagger$]}  & Type  \\ 
       & \multicolumn{1}{c}{RA}         & \multicolumn{1}{c}{Dec.}        &         &  & [mag] & \\ \hline
HCG~80  & \timeform{15h59m12.4s} & \timeform{+65\circ13'33.3''} & 0.02990 & &      & Group   \\
HCG~80a & \timeform{15h59m19.0s} & \timeform{+65\circ13'57.4''} & 0.02994 & $50.8\times10.8$& 15.66& Sd \\
HCG~80b & \timeform{15h59m21.5s} & \timeform{+65\circ13'22.7''} & 0.03197 & $20.4\times17.4$& 16.37& Sa \\
HCG~80c & \timeform{15h59m07.3s} & \timeform{+65\circ14'00.8''} & 0.03186 & $19.6\times16.4$& 16.06& Im \\
HCG~80d & \timeform{15h59m12.1s} & \timeform{+65\circ13'19.3''} & 0.03038 & $27.2\times10.8$& 17.01& Im \\ \hline
   \multicolumn{7}{@{}l@{}}{\hbox to 0pt{\parbox{180mm}{\footnotesize
      \par\noindent
      \footnotemark[*]Optical coordinates of the object from \citet{Hickson_etal_1989}.
     \par\noindent
      \footnotemark[$\dagger$] Major axis [arcsec]$\times$ minor axis [arcsec].
      \par\noindent
      \footnotemark[$\ddagger$] B magnitude.
     }\hss}}
\end{tabular}
\end{center}
\end{table*}
%%%%%%%%%%%%%%%%%%%%%%%%%%%%%%%%%%%%%%%%%%%%

\subsection{Source Detection in the ACIS-S3 Field}
\label{subsec:source_detection}

We searched for X-ray sources in the ACIS-S3 field of view with the
wavedetect algorithm with a significance parameter of $10^{-6}$ 
utilizing the 0.3--10 keV band image, and
detected 20 in total, including HCG~80a and HCG~80b. We also found
significant X-ray emission from the direction of HCG~80c. The counting
rates of four member galaxies are summarized in table~\ref{tab2}.  In
the following analysis, the detected point sources were excluded with a
radius of 7-times the size of the Point Spread Function (PSF),   
which is defined as the 40\% encircled energy radius at 1.5 keV at the
source position.

%%%%%%%%%%%%%%%%%%%%%%%%%%%%%%%%%%%%%%%%%%%%
\begin{table*}
\begin{center}
\caption{X-ray count rates, fluxes, and hardness ratios for member galaxies.}
\label{tab2}
\begin{tabular}{llllllll}\hline\hline
       & & & \multicolumn{2}{c}{Soft band} & \multicolumn{2}{c}{Hard band} & \\ \cline{4-5} \cline{6-7}
       & $R_{\rm spec}$\footnotemark[*] & $R_{\rm max}$\footnotemark[$\dagger$] & $S$\footnotemark[$\ddagger$] & $f_{\rm X,S}$\footnotemark[$\S$] & $H$\footnotemark[$\|$] & $f_{\rm X,H}$\footnotemark[$\#$] & $HR$\footnotemark[**]\\ \hline
HCG~80a          & 25 & 31 & $65.9\pm6.4$ & $3.4\pm 0.3$ 
                 & $5.8\pm3.5$ & $0.7\pm 0.4$ & $0.09\pm0.05$ \\
HCG~80a, nucleus & -- & 2 & $14.7\pm2.7$ & $0.7\pm 0.1$ 
                 & $3.5\pm1.3$ & $0.6\pm 0.2$ & $0.24\pm0.10$\\
HCG~80a, disk+halo & -- & $31-2$ & $50.3\pm5.8$ & $2.6\pm0.3$ 
                 & $2.9(<6.1)$ & $0.3(<0.7)$ & $0.06\pm0.06$ \\
HCG~80b, nucleus & 1.5 & 2 & $107.0\pm7.4$ & $5.3\pm 0.4$ 
                 & $34.9\pm4.2$ & $6.2\pm 0.7$ & $0.33\pm 0.05$ \\
HCG~80b, disk    & -- & $10-2$ & $6.2\pm1.9$ & $0.2\pm0.1$ 
                 & $0.4(<1.4)$ & $0.04(<0.15)$ & $0.06(<0.23)$ \\
HCG~80c & --   & 9.8   & $5.6\pm1.9$ & $0.2\pm 0.1$ 
                 & $0.3(<1.3)$ & $0.04(<0.15)$ & $0.06(<0.24)$\\
HCG~80d & --   & 13.6  & $1.7(<3.2)$ & $0.06(<0.13)$ 
                 & $<2.7$ & $<0.05$ & $1.6\pm1.4$\\ \hline
   \multicolumn{8}{@{}l@{}}{\hbox to 0pt{\parbox{180mm}{\footnotesize
      \par\noindent
      \footnotemark[*] The spectral extraction radius in arcsec (see subsection~\ref{subsec:member_spec}). 
      \par\noindent
      \footnotemark[$\dagger$] The maximum radius in arcsec, used to estimate the counting rate, 
      $S$ and $H$, and the hardness ratio, $HR$.
      \par\noindent
      \footnotemark[$\ddagger$] The X-ray counting rate in the 0.5--2 keV band, 
      $S~[10^{-4}{\rm counts\,s^{-1}}]$.
      \par\noindent      
      \footnotemark[$\S$] The X-ray flux in the 0.5--2 keV band, 
      $f_{\rm X,S}~[10^{-14}{\rm erg\,s^{-1}cm^{-2}}]$.
      \par\noindent            
      \footnotemark[$\|$] The X-ray counting rate in the 2--7 keV band, $H~[10^{-4}{\rm counts\,s^{-1}}]$.
      \par\noindent            
      \footnotemark[$\#$] The X-ray flux in the 2--7 keV band, 
      $f_{\rm X,H}~[10^{-14}{\rm erg\,s^{-1}cm^{-2}}]$.
      \par\noindent            
      \footnotemark[$**$] The hardness ratio, $HR=H/S$. 
      \par\noindent            
      The quoted errors are the $1\sigma$.
       }\hss}}
\end{tabular}
\end{center}
\end{table*}
%%%%%%%%%%%%%%%%%%%%%%%%%%%%%%%%%%%%%%%%%%%%

The X-ray maximum positions of HCG~80a and b in the ACIS-S3 image are
(\timeform{15h59m18.9s}, \timeform{+65\circ13'57.3''}) and
(\timeform{15h59m21.6s}, \timeform{+65\circ13'22.8''}), respectively.
Comparing these with the optical coordinates from
\citet{Hickson_etal_1989}, they are consistent within $\lesssim
1\arcsec$. However, due to the limited photon statistics of the
present X-ray data, there are uncertainties in the X-ray positions of
the galaxies. We thus assumed the optical positions from
\citet{Hickson_etal_1989} as their centers in our analysis.

\section{Analysis and Results}\label{sec:analysis_and_result}

We analyze the X-ray emission from the individual galaxies in
subsections~\ref{subsec:member_image}, \ref{subsec:hardness}, and
\ref{subsec:member_spec}. In subsection~\ref{subsec:diffuse_emission}
we exclude them from the group region and constrain the X-ray emission
from the hot intragroup gas.

\subsection{X-Ray Morphologies of Member Galaxies}
\label{subsec:member_image}

As shown in table~\ref{tab2}, because HCG~80a and b were detected with more
than $10\sigma$ significance, we performed detailed analyses of the
surface brightness distributions and energy spectra for those two
galaxies.  Since the significance of the X-ray emission from HCG~80c
and d are low ($2.6 \sigma$ and $\lesssim 2\sigma$ in the soft band,
respectively), we could only measure the fluxes (and luminosities
under the assumptions of the spectral models). The results of flux
estimations for the four member galaxies are also presented in
table~\ref{tab2}.

In order to constrain the spatial distribution of the X-ray emission
from HCG~80a and b, we investigated the surface brightness
distribution in two different ways: (1) a radial profile fitting and
(2) a 1-dimensional profile fitting. In the former analysis, we
compared the X-ray radial profile with the simulated PSF to constrain
the emission from a nuclear source and the presence of diffuse
emission. Since we found significant extended emission, particularly
from HCG~80a, we studied the spatial distribution in detail in a
latter analysis. We corrected the positional dependence of the
telescope and the detector responses with exposure maps.

\begin{figure}
  \begin{center}
    \FigureFile(80mm,80mm){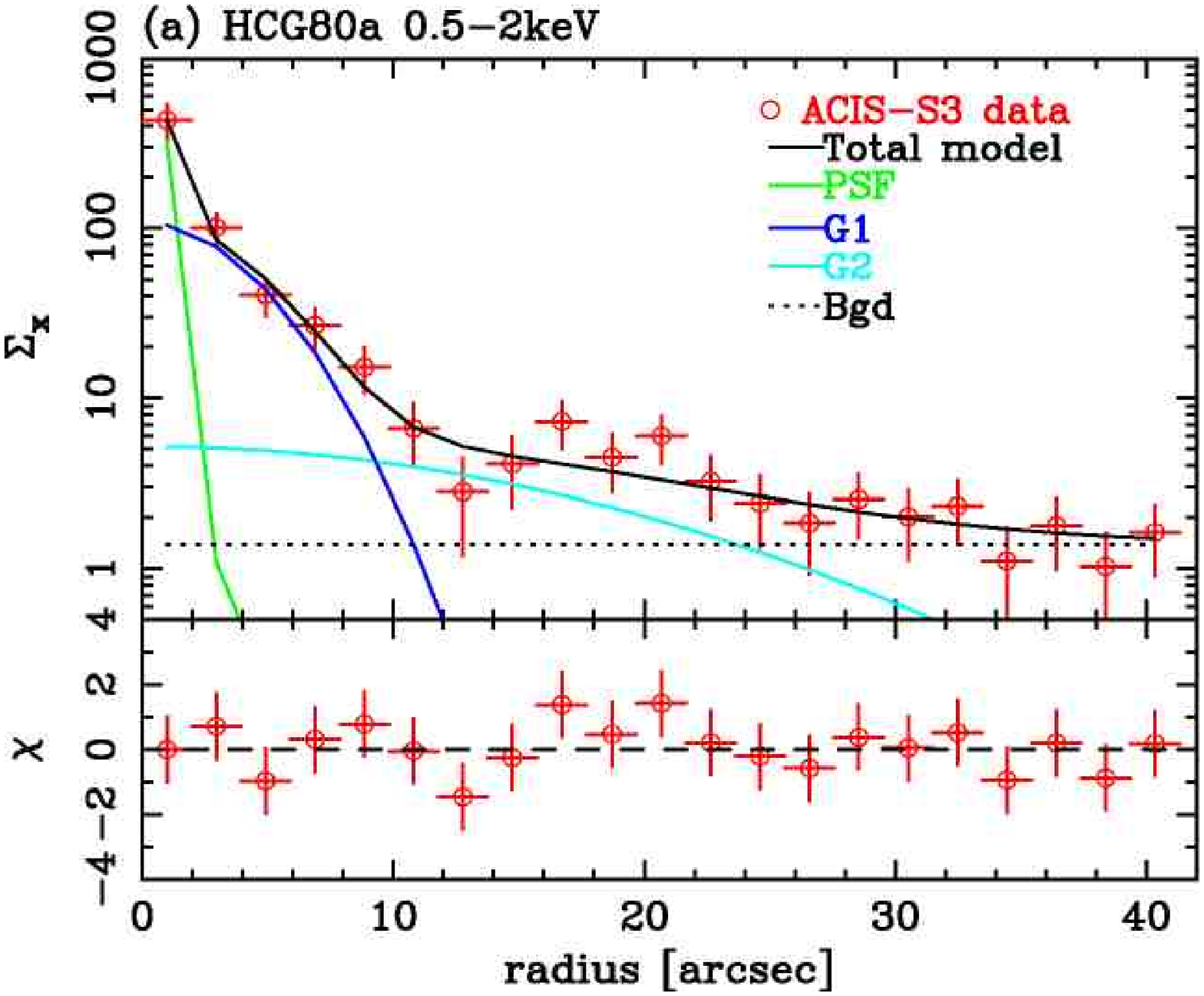}
    \FigureFile(80mm,80mm){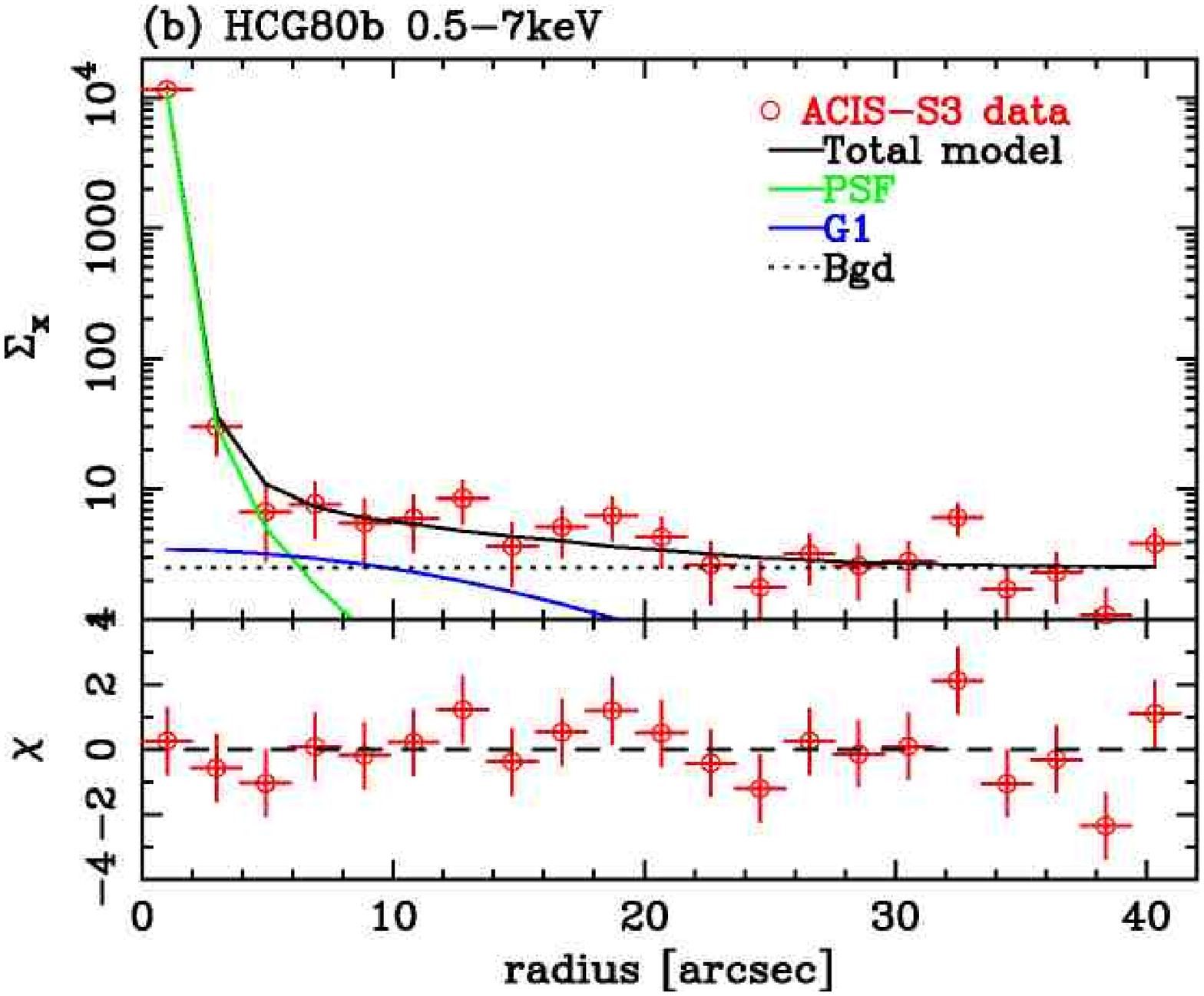}
  \end{center}
  \caption{Radial profile fitting for HCG~80a and b with the PSF and
    Gaussian(s) models.  In panels (a) and (b), the ACIS-S3 radial surface
    brightness distribution for HCG~80a in the 0.5--2 keV and HCG~80b in
    the 0.5--7 keV are indicated by red crosses.  The best-fit total
    model is shown with the black solid line in each panel, where the
    PSF and Gaussian components representing the disk and halo
    emissions are also indicated by the green, blue, and light-blue
    lines.  The background intensity is indicated by the horizontal
    dotted line.  In the upper panels $\Sigma_{\rm X}$ is in units of
    $10^{-9} {\rm photons\,s^{-1}cm^{-2}arcsec^{-2}}$, while in the
    bottom panels the residuals of the fits are shown in units of
    $\sigma$.  }\label{fig2}
\end{figure}

In the case of (1), we produced soft (0.5--2 keV) and hard (2--7 keV)
band images and calculated the radially averaged surface brightness
distributions centered at HCG~80a and b, as shown in
figure~\ref{fig2}.  We fitted them separately with the PSF models,
which were created by the HRMA PSF simulator, ChaRT, while specifying
the locations of the sources and the spectral models derived from the
spectral analysis (subsection~\ref{subsec:member_spec}). The PSF
models were generated with sufficient photons so that the statistical
uncertainty would be less than 10\%, thus negligible compared to the
Poisson error of the present data. We investigated the positional
dependence of the background intensity using both the blank-sky data
and the present data, to find that the background can be regarded as
being constant within the statistical errors. The background
intensities were then estimated from the $84\arcsec<r<160\arcsec$ ring
region of the present pointing data to be $(4.73\pm0.20)\times10^{-7}$
and $(5.31\pm0.21)\times10^{-7}$ counts s$^{-1}$ arcsec$^{-2}$ in the
soft and hard bands, respectively (the errors are $1\sigma$), and
included as fixed constants in the fits.

%%%%%%%%%%%%%%%%%%%%%%%%%%%%%%%%%%%%%%%%%%%%
%%%%%%%%%%%%%%%%%%%%%%%%%%%%%%%%%%%%%%%%%%%%

\subsubsection*{HCG~80a}\label{subsec:morphology_HCG80a}
For HCG~80a, there is clearly an extended emission, particularly in
the soft energy band compared with the PSF (figure~\ref{fig2}a).  We
then evaluated the emission using the Gaussian functions to find that
the single-component Gaussian model can not sufficiently fit the data,
suggesting that at least two components are necessary to describe
them.  Thus, in figure~\ref{fig2}a, we show the best-fit model
consisting of the PSF and two Gaussians, where the diffuse emission is
seen out to the maximum radius of $R_{\rm max}\sim 31\arcsec$ above
the $2\sigma$ background level. We analyze the significance of the
emission components and the properties in more detail later.

In the hard energy band, there is excess emission over the background
within $2\arcsec$ from the HCG~80a optical center. The 2--7 keV
counting rate is $(3.5\pm1.3)\times10^{-4}$ counts s$^{-1}$. Under the
current statistics, we could not constrain the spatial
distribution. However, taking into account the spectral hardness
inferred from the analysis in subsection~\ref{subsec:hardness} and the
nuclear activity reported by \citet{Shimada_etal_2000}, we suggest
that the hard emission may be attributed to the central AGN in HCG
80a.

\begin{figure}
  \begin{center}
   \FigureFile(80mm,80mm){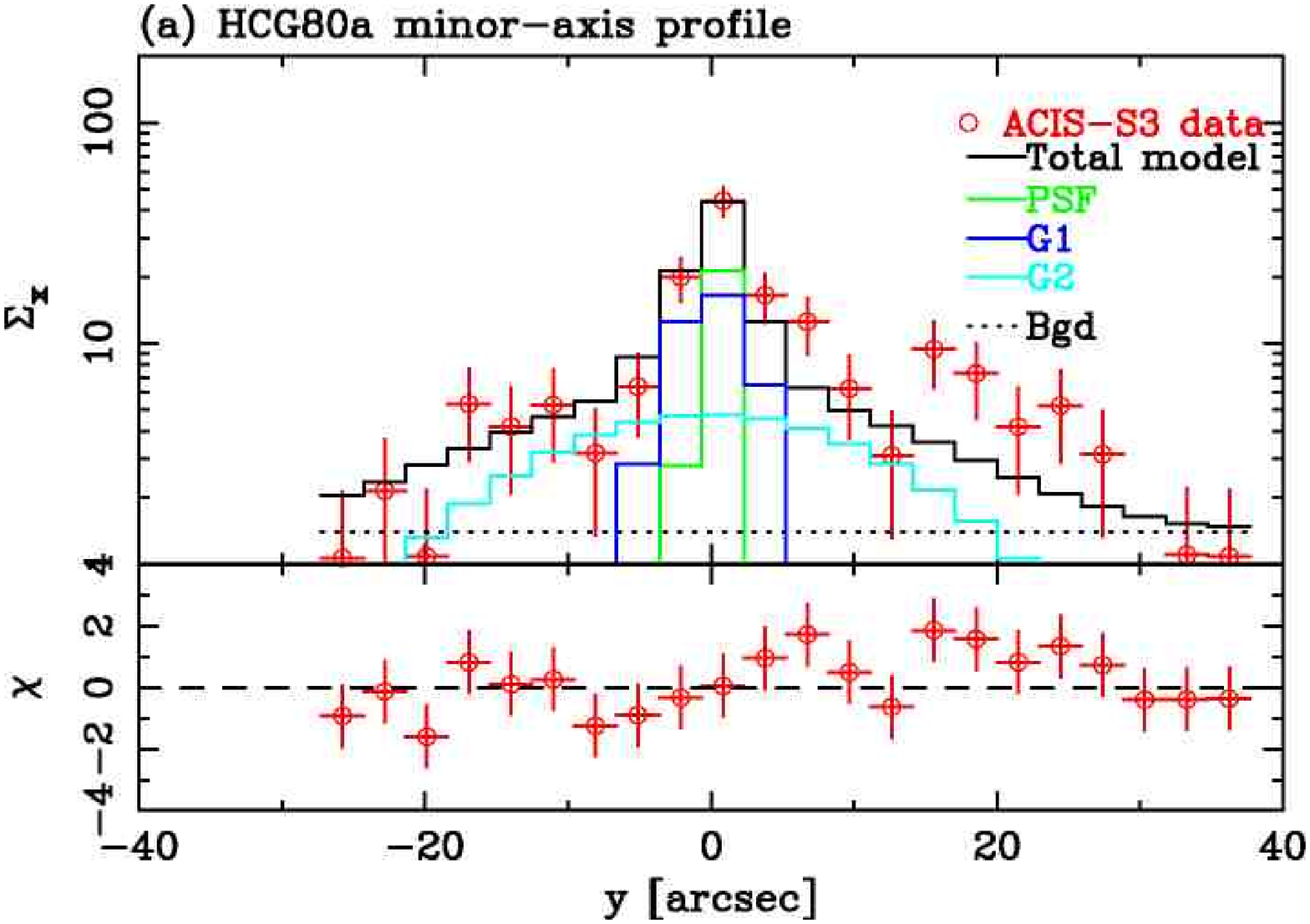}
   \FigureFile(80mm,80mm){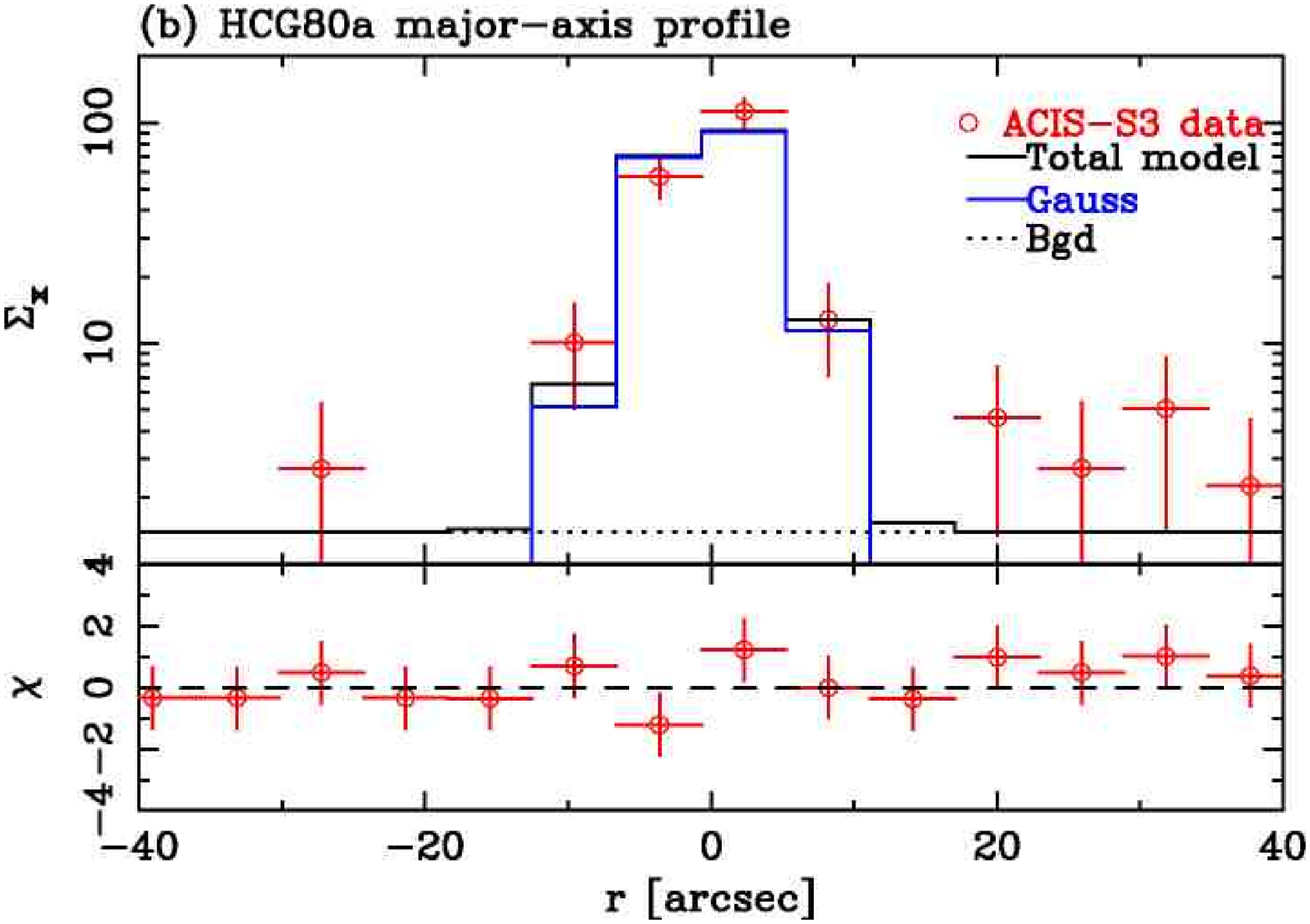}
  \end{center}
  \caption{Minor-axis and major-axis X-ray surface brightness
    distributions.  In the panels (a) and (b), the crosses denote the
    observed 1-dimensional surface brightness profiles accumulated
    within $|r|<24\arcsec$ of the major axis and $|y|< 5\arcsec$ of
    the minor axis in the soft band, respectively.  The surface
    brightness, $\Sigma_{\rm X}$, is in units of $10^{-9} {\rm
      photons\,s^{-1}cm^{-2}arcsec^{-2}}$.  In the panel (a), the
    solid black line shows the result of the fitting with Model 3. The PSF,
    narrow Gaussian, broad Gaussian, and background components are
    indicated by the green, blue, light-blue, and dotted black lines,
    respectively. In the panel (b),  the solid black, blue, and dotted lines show the 
    best-fit total model, Gaussian and the background components,
    respectively. }\label{fig3}
\end{figure}

In the next step, we consider the soft diffuse emission from HCG~80a,
based on a 1-dimensional profile fitting.  Since the galaxy has a
nearly edge-on inclination of $i=86^{\circ}$, following
\citet{Rubin_etal_1982} (see also \cite{Nishiura_etal_2000}), the
emission from the halo region is expected to be clearly resolved.
From figure~\ref{fig1}b, we find that the emission extends nearly
along the perpendicular direction from the galactic disk of HCG~80a,
whose position angle is $64^{\circ}$ \citep{Shimada_etal_2000}.  We
thus extracted the 1D surface brightness profile along the minor axis
of the galaxy, accumulated within $|r|<24\arcsec$ of the major axis
(figure~\ref{fig3}a).  The minor-axis profile was rebinned by a factor
of 6; thus, each bin is $3\arcsec$. We carried out a $\chi^2$ fitting
with some simple models consisting of the Gaussian components for the
diffuse emission and the PSF for the central emission, whose centers
are fixed at the HCG~80a center position.  As a result, neither the
single Gaussian nor the PSF model could provide an acceptable fit;
however, by adding another Gaussian component, the fit was
significantly improved at the 95\% level by the F-test. We show the
results for the cases of Model 1 [equation (\ref{eq:model1})] and
Model 2 [equation (\ref{eq:model2})] in table~\ref{tab3}.
\begin{eqnarray}
	\Sigma_{\rm X}
        &=& \Sigma_{\rm PSF} 
        + \Sigma_{\rm G1,0}e^{-(y-y_0)^2/\sigma_{\rm G1}^2}, \label{eq:model1}\\
  \Sigma_{\rm X} &=& 
        \Sigma_{\rm G1,0}e^{-(y-y_0)^2/\sigma_{\rm G1}^2}
        + \Sigma_{\rm G2,0}e^{-(y-y_0)^2/\sigma_{\rm G2}^2}, \label{eq:model2}\\
	\Sigma_{\rm X} &=& \Sigma_{\rm PSF} 
        +\Sigma_{\rm G1,0}e^{-(y-y_0)^2/\sigma_{\rm G1}^2}
        + \Sigma_{\rm G2,0}e^{-(y-y_0)^2/\sigma_{\rm G2}^2}, \label{eq:model3}
\end{eqnarray}
where $y_0$ is fixed at the center of HCG~80a, and we assume
$\sigma_{\rm G1}< \sigma_{\rm G2}$.  We also fit the profile with
Model 3 [equation (\ref{eq:model3})] and examine the significance of
the emission from the central point source.  Although the low photon
statistics in the central emission resulted in no significant
improvement over Model 2 in the F-test at the 95\% confidence, we
adopt Model 3 in the rest of our analysis regarding the presence of
the optically identified AGN \citep{Shimada_etal_2000} and the
observed high hardness ratio in subsection \ref{subsec:hardness}.
%%%%%%%%%%%%%%%%%%%%%%%%%%%%%%%%%%%%%%%%%%%%

%%%%%%%%%%%%%%%%%%%%%%%%%%%%%%%%%%%%%%%%%%%%
\begin{table*}
\begin{center}
\caption{Results of the 1D fitting for HCG~80a in the soft band.}
\label{tab3}
\begin{tabular}{llll}\hline\hline
	& Model 1 & Model 2 & Model 3 \\
Parameter	& PSF+G1     & G1+G2 & PSF+G1+G2 \\ \hline 
PSF [photons s$^{-1}$cm$^{-2}$] & 
$43_{-14}^{+13}\times 10^{-9}$ & -- & $25(<45)\times 10^{-9}$ \\
$\Sigma_{\rm G1,0}$ [photons s$^{-1}$cm$^{-2}$arcsec$^{-2}$]& 
$7.6_{-2.8}^{+5.2}\times 10^{-9}$ & 
$36.4_{-12.6}^{+19.5}\times 10^{-9}$ &  
$17.2_{-12.4}^{+31.7}\times 10^{-9}$ \\
$\sigma_{\rm G1}~[\rm arcsec]$
& $10\pm 4$ 
& $2\pm 1$ 
& $3_{-1}^{+2}$ \\
$\Sigma_{\rm G2,0}$ [photons s$^{-1}$cm$^{-2}$arcsec$^{-2}$]& -- & 
$5.6_{-2.8}^{+3.8}\times 10^{-9}$ & 
$4.7_{-2.8}^{+3.5}\times 10^{-9}$ \\ 
$\sigma_{\rm G2}~[\rm arcsec]$ 
&--  
& $12_{-4}^{+6}$ 
& $12_{-4}^{+8}$ \\
$\chi^2/{\rm dof}$ & 26.2/19 & 22.7/18 & 20.6/17 \\ \hline
 \end{tabular}
\end{center}
\end{table*}
%%%%%%%%%%%%%%%%%%%%%%%%%%%%%%%%%%%%%%%%%%%%

%--- AGN + disk + halo ---%
From the fitting with Model 3, the width for the narrow Gaussian
component was obtained to be $\sigma_{\rm G1}=3^{+2}_{-1}$~arcsec
($=1.8^{+1.2}_{-0.6}$~kpc), which is very consistent with the optical
scale of the galactic disk.  The broad Gaussian component has a width
of $\sigma_{\rm G2}=12^{+8}_{-4}$~arcsec ($=7.2^{+4.8}_{-2.4}$~kpc),
and thus largely extends compared to the disk scale.  Accordingly,
we refer to the narrow and broad Gaussian components, respectively,
as the ``disk'' and ``halo'' components hereafter.  The extent of the
halo emission detected above the $2\sigma$ background level is
$-26\arcsec<y<26\arcsec$, corresponding to 31.2 kpc in total.  The
origin of this huge X-ray halo is discussed in section
\ref{sec:discussion}.

We estimated the luminosity of each component assuming the MEKAL model
with an average temperature of 0.56~keV for the disk and the halo, and
a power-law model with $\Gamma=2.0$ for the nucleus. Here, the same
definitions as those of \citet{Strickland_etal_2004a} are used to
define the emission regions for the nuclear (N), disk (D), and halo(H)
components.  Namely the luminosities ($L_{\rm X,N}$, $L_{\rm X,D}$,
and $L_{\rm X,H}$) are accumulated from the $r<1$~kpc circle, a
rectangular aperture oriented along the minor axis of major-axis
length 28.9 kpc and extending between $ -2~{\rm kpc} < y < 2$~kpc, and
a $28.9~{\rm kpc}\times39.1~{\rm kpc}$ rectangular aperture oriented
along the minor axis with excluding the N and D regions.  We obtained
$L_{\rm X,D} = 2.2\times10^{40}$ erg s$^{-1}$, $L_{\rm X,H}
=3.8\times10^{40}$ erg s$^{-1}$, and $L_{\rm X,N}=1.7\times10^{40}$
erg s$^{-1}$ in the 0.5--2 keV, respectively.  The results are also
given in table~\ref{tab5}.

%--- North vs South --%
We also notice that the intensity of the halo emission from the
Northern hemisphere is stronger than that of the Southern by a factor
of $\sim 2.7\pm2.1$ if we fit the Northern and Southern halos
separately with Model 3 while fixing the model parameters, except for
$\Sigma_{\rm G2,0}$ at the best-fit values derived in the above
analysis.  Thus, there may be some asymmetry. However, because the
current statistics is limited, we could not further constrain the
spatial distribution of the halo emission.

We show the major-axis profile accumulated within $|y|< 5\arcsec$ of
the minor axis in figure~\ref{fig3}b. It could be fitted by a Gaussian
with $\sigma=4\pm 1$~arcsec ($=2.4\pm0.6$~kpc) and a normalization for
the central surface brightness of $108_{-27}^{+49}\times 10^{-9}~{\rm
  photons\,s^{-1}cm^{-2}arcsec^{-2}}$.  We found that the width is
close to that of the narrow component in Model 3, namely $\sigma\sim
\sigma_{\rm G1}$. Thus, the disk emission is concentrated within about
\timeform{3.5"} from the center, while we did not find significant
emission extending over the optical scale of the galactic disk.

\subsubsection*{HCG~80b}
From a comparison to the simulated PSFs, we found that the innermost
data points of the radial surface brightness distributions are well 
consistent with the PSFs in both the soft and hard energy bands, and
resultant $\chi^2$ values of 24.6 and 34.4 for 20 degrees of
freedom. However, we found that there are systematic residuals over
the PSFs in several consecutive bins around $r\sim10\arcsec$, which
can be attributed to the extra emission around the point source.
Because the emission is found within a radius roughly corresponding to
the size of the optical disk, $\sim 10\arcsec$, we refer to it as the
disk component, while referring to the central point source as the
nucleus component hereafter.  The photon counts coming from the disk
component in a radius range of $2\arcsec<r<10\arcsec$ are $12\pm4$ and
$<3$, in the soft and hard bands, respectively (see also
table~\ref{tab2}).

In order to further constrain the emission profile, we attempted to
fit the 0.5--7 keV radial profile with the PSF plus Gaussian model, as
shown in figure~\ref{fig2}.  This provided an acceptable fit with
$\chi^2/{\rm dof}=19.3/18$, and the PSF intensity was obtained as
$(1.13\pm0.17)\times10^{-5}~{\rm photons\,s^{-1}cm^{-2}}$; the
Gaussian normalization and the width are
$3.5^{+3.5}_{-2.6}\times10^{-9}~{\rm
  photons~s^{-1}cm^{-2}arcsec^{-2}}$ and $12^{+7}_{-5}$~arcsec
($=7^{+4}_{-3}$~kpc), respectively. Because the model parameters are
associated with large uncertainties, the emission profile,
particularly for the diffuse emission, is not well constrained, which
we consider as being due to the low signal-to-noise ratios at the
outer radius.  We therefore decide to fix the maximum radius at the
size of the optical disk, $R_{\rm max}=10\arcsec$, when evaluating the
X-ray intensities for the disk emission.  Within $2\arcsec$ from the
HCG~80b peak, the central nuclear component dominates the total
emission, and we use the maximum radius of $R_{\rm max}=2\arcsec$ for
the HCG~80b nucleus.  The luminosities of the nucleus ($r<2\arcsec$)
and the disk ($2\arcsec<r<10\arcsec$) were estimated to be
$(2.9\pm0.2)\times10^{41}$~erg s$^{-1}$ and
$(5.8\pm1.8)\times10^{39}$~erg s$^{-1}$ in the 0.5--7 keV band
assuming a power-law model with $\Gamma=1.9$ (see subsection
\ref{subsec:member_spec_b}) and the MEKAL model with $kT=0.5$~keV and
$Z=0.1$~solar, respectively.

\subsection{Hardness Ratio Analysis}\label{subsec:hardness}

To provide a quantitative evaluation of the spatial variation of the
X-ray spectra, the hardness ratios for the central and the surrounding
disk (+ halo) regions are given in table~\ref{tab2}. We define the
hardness ratio as $HR \equiv H/S$, where $S$ and $H$ are the photon
counts in the 0.5--2 and 2--7 keV bands, respectively.  Due to the low
significance of the emission from HCG~80c and d, we give the values
derived for the entire galaxy regions for these two galaxies.  There
is an indication that the HCG~80a central emission ($r<2\arcsec$) is
hard compared with the outer ($2\arcsec<r<31\arcsec$) region. If we
simply assume the spectra as being described by a power-law (or the
$Z=0.1$~solar MEKAL) model attenuated by Galactic absorption, the
power-law indices (or the gas temperature) for the nucleus and the
disk + halo regions of HCG~80a correspond to $\Gamma=2.2\pm 0.4$
($kT=2.6\pm 1.4$~keV) and $\Gamma=3.5(>2.8)$ ($kT=1.0 (<1.4)$~keV),
respectively.  While the HCG~80b nucleus region ($r<2\arcsec$) shows a
large $HR$ value, corresponding to a power-law index of $1.9\pm 0.2$,
the $HR$ for the outer disk region ($2\arcsec<r<10\arcsec$) was not
constrained. In the next subsection we explore the spectra from the
HCG~80a ``nucleus + disk + halo'' and the HCG~80b ``nucleus'' regions
based on thermal and/or non-thermal spectral modeling to constrain
their origins.

\citet{Strickland_etal_2004a} suggested, based on the Chandra data for
ten star-forming galaxies, that the disk emission tends to be harder
than the halo emission; however, we did not find any meaningful HR
variation between the two regions in HCG~80a due to the poor photon
statistics of the present data.

\subsection{X-Ray Spectra of Member Galaxies}
\label{subsec:member_spec}

\begin{figure}
  \begin{center}
    \FigureFile(80mm,80mm){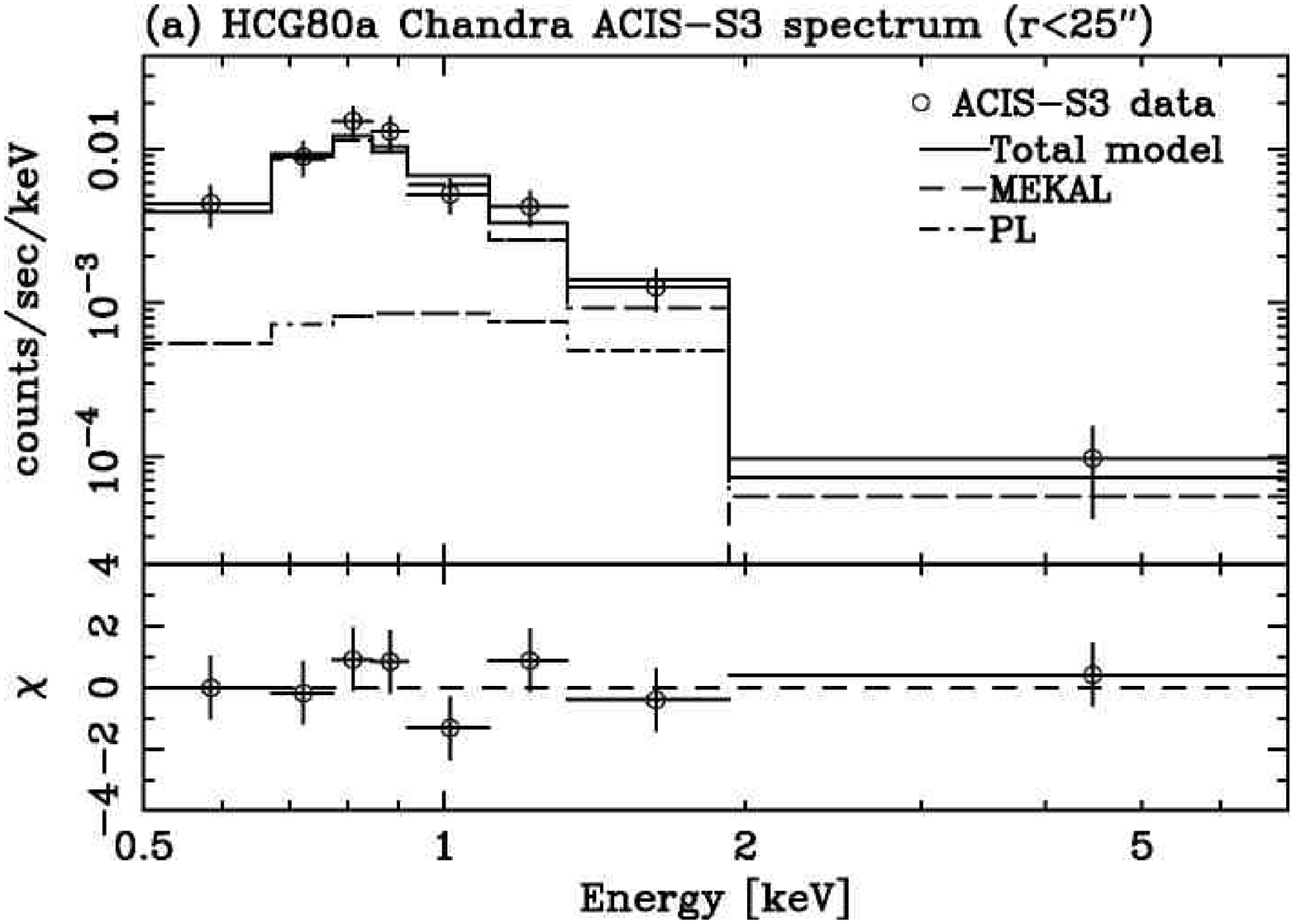}
    \FigureFile(80mm,80mm){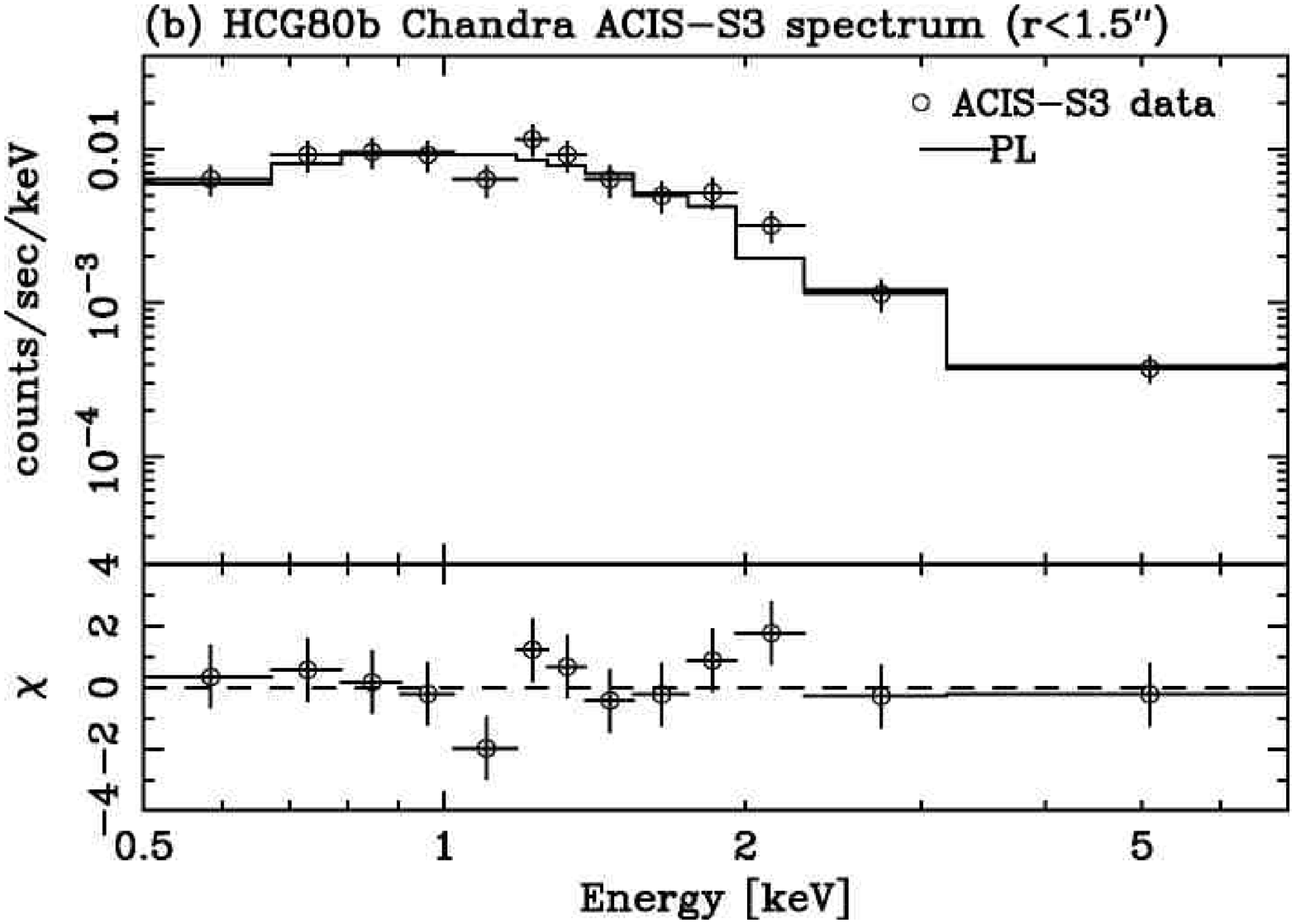}
  \end{center}
  \caption{Chandra ACIS-S3 spectra of HCG~80a (a) and HCG~80b (b).  In
    the upper panels, the open circles denote the observed spectra and
    the step functions show the best-fit spectral models (the
    MEKAL+Power-law model for HCG~80a and the Power-law model for
    HCG~80b) convolved with the telescope and the detector response
    functions.  In the lower panels, the residuals of the fit are
    shown.  }
  \label{fig4}
\end{figure}

We extracted X-ray spectra for HCG~80a and b from circular regions
with radii $R_{\rm spec}=25\arcsec$ and \timeform{1.5"}, respectively
(figure~\ref{fig4}). Note that $R_{\rm spec}$ was so chosen that about
90\% of the X-ray photons from each galaxy could be accumulated and
that two galaxies might not overlap each other. We fitted the spectra
with the power-law model and the MEKAL thin-thermal plasma model
(\cite{Mewe_etal_1985}, 1986;
%\cite{Mewe_etal_1986}; 
\cite{Kaastra_1992}; \cite{Liedahl_etal_1995}). The absorption
column density was fixed at the Galactic value, $N_{\rm H} = 2.5\times
10^{20}~{\rm cm^{-2}}$ \citep{Dickey_Lockman_1990}.

%%%%%%%%%%%%%%%%%%%%%%%%%%%%%%%%%%%%%%%%%%%%
%%%%%%%%%%%%%%%%%%%%%%%%%%%%%%%%%%%%%%%%%%%%

\subsubsection*{HCG~80a}\label{subsec:member_spec_a}
For HCG~80a, the power-law spectral model was rejected at the 99\%
confidence level. On the other hand, the MEKAL model provided a good
fit to the data with a $\chi^2$ value of 5.9 for 5 degrees of
freedom. We obtained the temperature and the metallicity to be
$0.59^{+0.12}_{-0.10}$ keV and $0.07^{+0.18}_{-0.05}$ solar,
respectively. Though the AGN emission was estimated to be only $\sim
20$\% of the total emission from the image analysis, we checked
whether the temperature determination of the intragroup gas was
influenced by the AGN component in the following two ways: (1)
excluded the central $r<5\arcsec$ circular region from the HCG~80a
overall spectrum, and fit it with the MEKAL model, and (2) fit the
overall spectrum with the two-component model, where the power-law
index and the metallicity were, respectively, fixed at 2.0 and 0.1
solar. We found that both of the analyses yielded consistent
temperatures ($\sim 0.6$~keV) within the errors. We give the results
for case (2) in table~\ref{tab4}. We also confirmed that the ratio of
the X-ray luminosities between the AGN and the diffuse components is
consistent with the result of the image analysis under Model 3 within
10\%.

%%%%%%%%%%%%%%%%%%%%%%%%%%%%%%%%%%%%%%%%%%%%
\begin{table*}
\begin{center}
\caption{Results of spectral fittings for HCG~80a and HCG~80b.}
\label{tab4}
\begin{tabular}{lllll}\hline\hline
Galaxy & Model     & Parameter & Value (90\% error) & $\chi^2/{\rm dof}$ \\ \hline
HCG~80a & MEKAL     & $kT$~[keV] & $0.59^{+0.12}_{-0.10}$  & \\
($R_{\rm spec}=25\arcsec$)& & $Z$~[solar]& $0.07^{+0.18}_{-0.05}$  & \\
       &           & $z$       & $0.02994$ (F)& \\
       &           & $k_{\rm M}$\footnotemark[*]& $9.9^{+8.3}_{-5.7}\times10^{-5}$  & 5.9/5\\ \cline{2-5}
       & PL +MEKAL & $\Gamma$  & 2.0 (F) & \\ 
       &           & $k_{\rm P}$\footnotemark[$\dagger$] & $<5.2\times10^{-6}$  & \\ 
       &           & $kT$~[keV]  & $0.56^{+0.12}_{-0.17}$  & \\
       &           & $Z$~[solar] & 0.1 (F)  & \\
       &           & $z$       &0.02994 (F)& \\
       &           & $k_{\rm M}$\footnotemark[*] & $7.4^{+2.9}_{-1.9}\times10^{-5}$  & 4.4/5\\ \cline{3-5}
       &           & $f_{\rm X,P}$ [erg s$^{-1}$cm$^{-2}$]\footnotemark[$\ddagger$]  & $<2.1\times10^{-14}$ & \\
       &           & $L_{\rm X,P}$ [erg s$^{-1}$]\footnotemark[$\ddagger$] & $<4.6\times10^{40}$ & \\ 
       &           & $L_{\rm bol,P}$ [erg s$^{-1}$]\footnotemark[$\ddagger$]  & $<6.1\times10^{40}$ & \\ 
       &           & $f_{\rm X,M}$ [erg s$^{-1}$cm$^{-2}$]\footnotemark[$\S$]  & $(2.8\pm 0.5)\times10^{-14}$& \\
       &           & $L_{\rm X,M}$ [erg s$^{-1}$]\footnotemark[$\S$]  & $6.5^{+0.8}_{-1.2}\times10^{40}$ & \\ 
       &           & $L_{\rm bol,M}$ [erg s$^{-1}$]\footnotemark[$\S$]  & $7.8^{+1.2}_{-1.6}\times10^{40}$ & \\ \hline
HCG~80b & PL & $\Gamma$  & $1.94^{+0.19}_{-0.18}$ & \\
($R_{\rm spec}=\timeform{1.5"}$)&           & $k_{\rm P}$\footnotemark[$\dagger$] & $2.5^{+0.3}_{-0.3}\times10^{-5}$ & 10.7/11\\ \cline{3-5}
       &           & $f_{\rm X,P}$ [erg s$^{-1}$cm$^{-2}$]\footnotemark[$\ddagger$] & $10.6^{+1.6}_{-1.8}\times10^{-14}$ & \\
       &           & $L_{\rm X,P}$ [erg s$^{-1}$]\footnotemark[$\ddagger$]  & $(2.6\pm 0.3)\times10^{41}$& \\ 
       &           & $L_{\rm bol,P}$ [erg s$^{-1}$]\footnotemark[$\ddagger$]  & $(3.5\pm 0.4)\times10^{41}$& \\ \cline{2-5}
       & MEKAL     & $kT$~[keV]& $3.5^{+1.6}_{-1.0}$  & \\
       &           & $Z$~[solar]& $<0.17$  & \\
       &           & $z$       &0.03197 (F)& \\
       &           & $k_{\rm M}$\footnotemark[*]  & $1.07^{+0.16}_{-0.14}\times10^{-4}$  & 12.7/10\\ \hline
      \multicolumn{5}{@{}l@{}}{\hbox to 0pt{\parbox{180mm}{\footnotesize
      \par\noindent
       \footnotemark[*]  Normalization factor for the MEKAL model, $k_{\rm M}=\int n_{\rm e} n_{\rm H}
         dV/4\pi (D_{\rm A} (1+z))^2~[10^{-14}~{\rm cm^{-5}}]$, where $D_{\rm A}$ is the
         angular diameter distance to the source. 
      \par\noindent         
       \footnotemark[$\dagger$]  Normalization factor for the power-law (PL) model, $k_{\rm P}~{\rm
         [photons\,keV^{-1}cm^{-2}s^{-1}]}$ at 1keV. 
      \par\noindent                  
         (F) Fixed parameters.
      \par\noindent                            
       \footnotemark[$\ddagger$]  The 0.5--7 keV X-ray flux and the 0.5--7 keV luminosity for the PL model.
      \par\noindent                                   
       \footnotemark[$\S$]  The 0.5--7 keV X-ray flux and the 0.5--7 keV luminosity for the MEKAL model.
       }\hss}}
\end{tabular}
\end{center}
\end{table*}
%%%%%%%%%%%%%%%%%%%%%%%%%%%%%%%%%%%%%%%%%%%%

\subsubsection*{HCG~80b}\label{subsec:member_spec_b}
We found that the HCG~80b spectrum can be fitted either by the
power-law model or the MEKAL model.  Even though the MEKAL temperature
of $kT\sim 3.5$~keV may be consistent with a collection of Low Mass
X-ray Binaries, the observed luminosity of $2.6\times10^{41}$~erg
s$^{-1}$ is by more than 3 orders higher than the values for normal
spiral galaxies.  On the other hand, the power-law index of
$\Gamma=1.94^{+0.19}_{-0.18}$, deduced from the X-ray spectral fitting
is consistent with the values of known AGNs.  We also estimated the
$\alpha_{\rm OX}$ index, which is the slope of a hypothetical power
law connecting the B band and 2 keV, to be 1.9.  The value is larger
compared with the result of the ROSAT large quasar survey, but within
the scatter of the quasars (figure 2 of \cite{Green_etal_1995}).
Thus, the emission from the central $r<\timeform{1.5"}$ region is most
likely to originate from the AGN in the galaxy.

\subsection{Constraints on the Hot Diffuse Emission}\label{subsec:diffuse}
\label{subsec:diffuse_emission}
In order to constrain the X-ray emission from the hot intragroup
medium, we defined the group region with a $r=84\arcsec=50.5~{\rm
  kpc}$ circle, whose center is the same as that of HCG~80,
\timeform{15h59m12.4s}, \timeform{+65\circ13'33.3''}
(Hickson et al. 1989), 
%\citep{Hickson_etal_1989}, 
which encompasses the optical disks of the four member galaxies
(figure~\ref{fig1}b). We subtracted the background and the galaxy
contributions from the total photon counts in the group region and
derived the intensity of the intragroup emission. Note that we used
the background intensity estimated from the outer-ring region. The
source counts within $31\arcsec$ from the brightest member, HCG~80a,
were estimated based on the results of the image analysis presented in
subsection~\ref{subsec:morphology_HCG80a}, while for HCG~80b--d, the
observed source counts within circles of radii $R_{\rm
  max}=10\arcsec$, \timeform{9.8"} and \timeform{13.6"}, 
approximately equal to the sizes of the optical disks, were used.

\begin{figure}
  \begin{center}
    \FigureFile(80mm,80mm){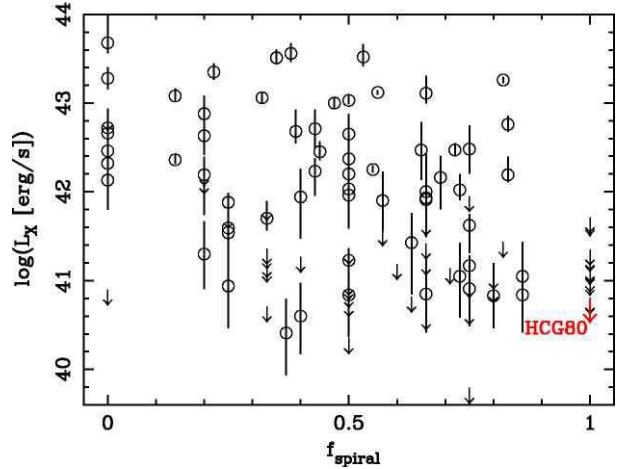}
  \end{center}
  \caption{$L_{\rm X}-f_{\rm spiral}$ relation.  The groups with
    constrained/unconstrained bolometric luminosity from the ROSAT
    PSPC survey \citep{Mulchaey_etal_2003} are plotted with open
    circle/arrow, as a function of the spiral fraction.  The upper
    limit on the bolometric luminosity for the hot diffuse gas in
    HCG~80, obtained from the present Chandra analysis, is indicated by the
    red arrow.  }\label{fig5}
\end{figure}

Subtracting the galaxy contributions, $130\pm12$, $223\pm15$,
$11\pm3$, $3(<5)$ counts for HCG~80a--d and the background, $213\pm 9$
counts from the total photon counts of $576\pm24$ in the group region,
we found that there is no significant emission from the hot IGM, and
that the $3\sigma$ upper limit is obtained to be 92 photons in the
0.5--2 keV band.  We also confirmed that the present estimation of the
IGM emission is not affected by the choice of the extraction radii,
$R_{\rm max}$ for the galaxies b--d; if we change $R_{\rm max}$ by a
factor of $0.5 - 1.5$, the result changes by only $\lesssim10$\%.  If
we further assume the temperature and the metallicity of the gas to be
comparable to those derived for HCG~16, $kT\sim0.5$ keV and $Z\sim0.1$
solar \citep{Belsole_etal_2003}, the bolometric luminosity is
constrained as $L_{\rm X} < 6.3\times10^{40}$ erg s$^{-1}$
($3\sigma$). We show the location of HCG~80 on the $L_{\rm X}-f_{\rm
  spiral}$ plane in figure~\ref{fig5}.

%%%%%%%%%%%%%%%%%%%%%%%%%%%%%%%%%%%%%%%%%%%%
%%%%%%%%%%%%%%%%%%%%%%%%%%%%%%%%%%%%%%%%%%%%

\section{Discussion}\label{sec:discussion}

From a high-resolution Chandra observation of the spiral-only group
HCG~80, we detected significant X-ray emission from three of the four
member galaxies (HCG~80a, b, and c), and investigated the spatial
distribution and the spectral features for HCG~80a and b in detail. In
particular, we discovered halo emission from HCG~80a, which extends to
$\sim 30$~kpc perpendicular to the galactic disk. We compare the X-ray
luminosity of the member galaxies to those of the optical in
subsection~\ref{subsec:discuss_lxlb}, and further discuss the origin
of the halo emission from HCG~80a in
subsection~\ref{subsec:discuss_starburst}.  On the other hand, we
found that there is no significant emission from the group region.
The absence of strong X-ray emission is discussed in
subsection~\ref{subsec:discuss_diffuse}.

\subsection{X-Ray Properties of the Member Galaxies}
\label{subsec:discuss_lxlb}

\begin{figure}
  \begin{center}
    \FigureFile(80mm,80mm){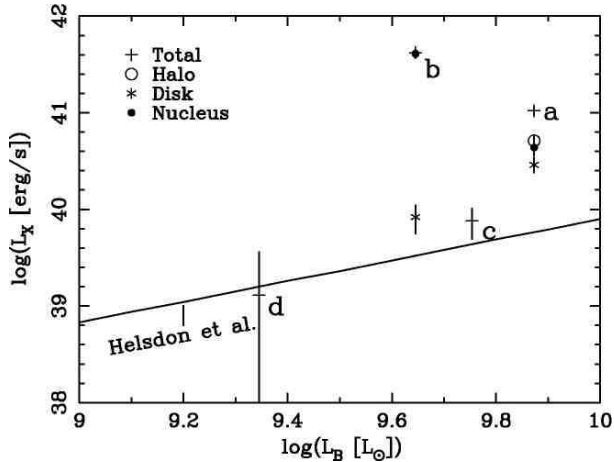}
  \end{center}
  \caption{$L_{\rm X}-L_B$ relation. The bolometric luminosity of the
    overall galaxy region is plotted against the B luminosity [\LO]
    for each galaxy (cross). For HCG~80a, the nucleus, disk and halo
    components are also separately indicated by the solid circle,
    asterisk, and open circle, respectively. For HCG~80b, the nucleus
    and the disk components are indicated by the solid circle and
    asterisk, respectively. The error bars are the $1\sigma$. The
    solid line represents the best-fit $L_{\rm X}-L_B$ relation for
    the late-type group galaxies \citep{Helsdon_etal_2001}. }\label{fig6}
\end{figure}

We show the relation between the bolometric X-ray luminosity, $L_{\rm
  X}$, and the $B$ band luminosity, $L_B$, for the HCG~80 members in
figure~\ref{fig6}. In addition to the overall luminosities of the
galaxies, the luminosities of the nucleus, disk, and halo regions are
separately plotted for HCG~80a and b. The X-ray luminosities were
estimated from the spectral analyses for HCG~80a and the HCG~80b
nucleus. However, because the spectra for the HCG~80b disk, HCG~80c
and d were not constrained, we assume the thermal emission with
$kT=0.5$~keV and $Z=0.1$~solar to convert the photon counting rates in
the soft band (table~\ref{tab2}) to $L_{\rm X}$. This assumption may
be valid because the hardness ratios listed in table~\ref{tab2} are in
agreement with the values for soft thermal emission.

%%%%%%%%%%%%%%%%%%%%%%%%%%%%%%%%%%%%%%%%%%%%
%%%%%%%%%%%%%%%%%%%%%%%%%%%%%%%%%%%%%%%%%%%%

Comparing with the best-fit $L_{\rm X}-L_B$ relation for the late-type
galaxies in the \citet{Helsdon_etal_2001} sample, we found that the
disk component of HCG~80b and the overall luminosities of HCG~80c and
d are consistent with their relation if we take into account the
measurement errors and the large data scatter among the other
late-type galaxies.  On the other hand, HCG~80a and b clearly show a
higher $L_{\rm X}$ value than expected from the $L_{\rm X}-L_B$
relation from \citet{Helsdon_etal_2001}. They noted that the galaxies
associated with the AGN and/or the starburst activities tend to lie
above the best-fitting line, namely they have enhanced X-ray emission
relative to the optical band. Thus considering from the thermal nature
of the extended emission of HCG~80a and the hard spectrum of the
point-like emission from HCG~80b found in the previous section, the
higher $L_{\rm X}$ values agree with their indication.

%*** LX -- LFIR relation ***
Furthermore, starburst galaxies are often identified based on their
high IR luminosity and warm FIR colors, $f_{60}/f_{100}>0.4$. The FIR
fluxes are available only for HCG~80a. From the IRAS 12-, 25-, 60-,
and 100-$\mu{\rm m}$ fluxes, $(f_{12},
f_{25},f_{60},f_{100})=(0.10,0.16,2.31,5.16)$~Jy, the IR luminosity is
$L_{IR}=12.3\times10^{10}~\LO$ utilizing $L_{\rm IR}=
5.67\times10^{5}D_{\rm
  Mpc}^2(13.48f_{12}+5.16f_{25}+2.58f_{60}+f_{100})\LO$
\citep{Sanders_Mirabel_1996}. This is consistent with the
$L_{\rm X}-L_{\rm IR}$ correlation for starburst galaxies
(\cite{Strickland_etal_2004b,
  Helsdon_etal_2001}). $f_{60}/f_{100}=0.45$; hence HCG~80a is ``FIR
warm''.  Thus, the above facts strongly support that the X-ray
emission from HCG~80a is produced by starburst activity.  In the next
subsection, we derive some physical parameters to characterize a 
starburst, and compare them to the previous measurements on other
starburst galaxies, in light of energy feedback from massive stars.

\subsection{Starburst Activity in HCG~80a}
\label{subsec:discuss_starburst}

The huge extraplaner emission discovered in HCG~80a reminds us of the
bipolar outflow in a bright starburst galaxy, such as M~82 (e.g.,
\cite{Lehnert_etal_1999}).  The X-ray luminosity from the HCG~80a halo
is determined to be $\sim 4\times10^{40}$ erg s$^{-1}$
(table~\ref{tab5}), which is even larger than that reported for bright
starburst galaxies.  We show a detailed comparison of the luminosities
derived in subsection~\ref{subsec:morphology_HCG80a} and the
emission-weighted X-ray temperature to the values reported for M~82
and NGC~253 \citep{Strickland_etal_2004a}.  We also show the
star-formation rate estimated from the IR luminosity, ${\rm
  SFR_{IR}}=21.1~\MO{\rm yr^{-1}}$, where ${\rm
  SFR_{IR}}=4.5\times10^{-44}L_{\rm IR}~{\rm [erg\,s^{-1}]}$
\citep{Kennicutt_1998}.
  
  %%%%%%%%%%%%%%%%%%%%%%%%%%%%%%%%%%%%%%%%%%%%
\begin{table*}
\begin{center}
\caption{Halo, disk, nuclear, and total luminosities.}
\label{tab5}
\begin{tabular}{llllllllll}\hline\hline
Galaxy & $\langle kT\rangle$\footnotemark[*]  & Band\footnotemark[$\dagger$]  & $L_{\rm X,H}$\footnotemark[$\ddagger$] 
& $L_{\rm X,D}$\footnotemark[$\S$]  & $L_{\rm X,N}$\footnotemark[$\|$]  
& $L_{\rm X,tot}$\footnotemark[$\#$]  
& $L_{B}$\footnotemark[**]   & $L_{\rm IR}$\footnotemark[**] 
& ${\rm SFR_{IR}}$\footnotemark[$\dagger\dagger$]  \\ \hline
HCG~80a & $0.56^{+0.12}_{-0.17}$ & 0.5--2 
       & $3.5\pm 0.5$ & $2.0\pm 0.4$ 
       & $1.6\pm 0.3$     & $7.1\pm 0.7$ & 0.75 & 12.3 & 21.1 \\ 
HCG~80a & $0.56^{+0.12}_{-0.17}$ & 0.3--2 
       & $4.5\pm 0.7$     & $2.6\pm 0.4$ 
       & $2.1\pm 0.4$     & $9.2\pm 0.9$ &  0.75 & 12.3 &  21.1 \\ \hline
M~82\footnotemark[$\ddagger\ddagger$]    & 0.37 & 0.3--2 & 0.41& 1.6 
       & 2.3 &4.3  & 0.33 & 5.36 & 9.2 \\
NGC~253\footnotemark[$\ddagger\ddagger$] & 0.25 & 0.3--2 & 0.12 & 0.33 
       & 0.094 & 0.55 & 0.58 & 2.10 & 3.6 \\ \hline
  \multicolumn{10}{@{}l@{}}{\hbox to 0pt{\parbox{180mm}{\footnotesize
      \par\noindent
      \footnotemark[*]  Emission-weighted X-ray temperature of the diffuse emission in keV. 
      \par\noindent
      \footnotemark[$\dagger$]  Energy band in keV used to calculate the X-ray luminosities.
      \par\noindent
      \footnotemark[$\ddagger,\S,\|$] X-ray luminosities 
      for the halo (H), the disk (D), and the nucleus (N) regions in $10^{40}$erg s$^{-1}$. 
      \par\noindent
      \footnotemark[$\#$]  Total X-ray luminosity for the H +
      D + N regions in $10^{40}$erg s$^{-1}$. 
      The errors of the luminosity measurements for HCG~80a are the $1\sigma$. 
      \par\noindent
      \footnotemark[**] $B$ and IR luminosities in $10^{10}$~\LO.  
      \par\noindent
      \footnotemark[$\dagger\dagger$] Star-formation rate from the IR luminosity,   
      ${\rm SFR_{IR}}=4.5\times10^{-44}L_{\rm IR}~{\rm [erg\,s^{-1}]}$ \citep{Kennicutt_1998}.
      \par\noindent
      \footnotemark[$\ddagger\ddagger$] The values were taken from \citet{Strickland_etal_2004a}.
       }\hss}}
\end{tabular}
\end{center}
\end{table*}
%%%%%%%%%%%%%%%%%%%%%%%%%%%%%%%%%%%%%%%%%%%%

In order to investigate the physical properties of the X-ray emitting
thermal plasma in the halo, we treated the emission region as a
cylinder of diameter 4 kpc and height 30 kpc, and estimated the
electron density and the gas mass. Thus, the following values are
meaningful only as order-of-magnitude estimates. The volume of the
cylinder is $V=1.1\times10^{67}~{\rm cm^{3}}$ and the emission
integral of the corresponding region is $EI=n_{\rm e}n_{\rm H} V
=1.44\times10^{64}~{\rm cm^{-3}}$ based on the result of the spectral
fitting. These yielded the electron density, the thermal pressure, the
thermal energy, and the gas mass for an average temperature of
$T=6.5\times10^{6}$~K, as follows:
\begin{eqnarray}
n_{\rm e}    & = & 4.0\times10^{-2}~{\rm cm^{-3}},\label{eq:halo_density} \\
P_{\rm th}/k & = & n_{\rm e} T = 2.6\times10^{5}~{\rm K\,cm^{-3}},\\
M_{\rm gas}  & = & 
\mu_{\rm e} m_{\rm p} n_{\rm e} V
= 4.3\times10^{8}~\MO,\\
E_{\rm th}   & = & 
\frac{3}{2}(n_{\rm e} + n_{\rm H}) kT V 
= 1.1\times10^{57}~{\rm erg},\label{eq:E_th}
\end{eqnarray}
where we adopt $n_{\rm H} = (\mu_{\rm e}/\mu_{\rm H})n_{\rm e}$,
$\mu_{\rm e}=1.167$, and $\mu_{\rm H}=1.40$. The above values are
higher by about a factor of $\gtrsim5$ than those obtained for NGC~253
\citep{Strickland_etal_2002} if we neglect any systematic error.

The radiative cooling timescale of the gas and the mass-cooling rate
are estimated to be $t_{\rm cool}\sim E_{\rm th}/L_{\rm bol} =
390$~Myr and $\dot{M}_{\rm gas} \sim 1.1~\MO{\rm yr^{-1}}$. Then,
supposing that the flow velocity may be approximated by the sound
speed of the gas, $v_{\rm flow}\sim v_{\rm s} = 292$~km s$^{-1}$, the
time necessary to travel the distance of 15~kpc (a half of the hight
of the cylinder) is $t_{\rm flow}\sim 50$~Myr. Since $t_{\rm cool} \gg
t_{\rm flow}$, the condition for maintaining halo emission seems to be
satisfied.

The above calculations and the high SF rate inferred from the IR
luminosity indicate that an enormous thermal energy of $\sim
10^{57}$~erg would be supplied through successive SN explosions and
the formation of superbubbles.  We will estimate the SN rate and also
compare the estimated quantities with the ``disk blowout'' condition
to test the plausibility of the present interpretation.

Assuming a Type II supernova energy input of $10^{51}$~erg and a
canonical value for the thermalization efficiency of 10\%, the thermal
energy contained in the hot gas requires $\sim 10^{7}$ SNe.  Thus, the
SN rate is expected to be $\sim 10^{7}/t_{\rm flow} = 0.2~{\rm
  yr^{-1}}$ to keep the $\sim 10^{7}$ ejecta in the halo region.
Alternatively, with the IR luminosity and the relation from
\citet{Heckman_etal_1990}, the SN rate is $R_{\rm SN} = 0.2L_{\rm
  IR}/10^{11}\LO\sim 0.25~{\rm yr^{-1}}$.  Then, if the successive
star formation had lasted in the past for $t_{\rm SF}=40$~Myr, and one
supernova may supply thermal energy of $10^{50}$~erg s$^{-1}$, the SN
rate of $0.25~{\rm yr^{-1}}$ can account for the thermal energy in the
halo of $E_{\rm th}\sim 10^{57}$~ergs [equation~(\ref{eq:E_th})].  The
duration of $t_{\rm SF}=40$ Myr is comparable to the timescale of the
outflow, $t_{\rm flow}$, and seems also to be reasonable from the
point of view of the typical lifetime of massive stars, $\sim 10$ Myr.

We next consider whether the gas can really escape from the galaxy
potential well against the gravitational force.  The escape velocity
was estimated to be $v_{\rm esc}\sim 120$~km s$^{-1}$ (or equivalently
$kT\sim 0.14$~keV) utilizing a mass-to-light ratio of $M/L_B\simeq
60h(R/0.1{\rm Mpc})\MO/\LO$ for spiral galaxies
\citep{Bahcall_etal_1995}. Here, we assumed that the galaxy mass
within $R$ is given by $M\sim Rv_{\rm esc}^2/G$, and adopted
$R=15$~kpc, which corresponds to the isophotal radius, $R_{25}$
\citep{Hickson_1993}.  Note that the rotation curve was measured
within the central $r\lesssim4$~kpc by \citet{Nishiura_etal_2000}.
Though it is difficult to infer $v_{\rm esc}$ from their result, due
to the existence of asymmetry between the approaching and receding
sides of the galaxy, the average rotation velocity is roughly $\sim
130$~km s$^{-1}$, and thus comparable to the value estimated above.
Therefore, the observed temperature of 0.6~keV is sufficiently high
for the gas to escape into intergalactic space.

The critical mechanical luminosity for the disk blowout [see
  \citet{Strickland_etal_2004b} and references therein] is calculated
as $L_{\rm crit}=4.2\times10^{40}$~erg s$^{-1}$. The mechanical energy
injection of the halo may be given by $L_{\rm W}\sim L_{\rm X,
  H}=3.5\times10^{40}$~erg s$^{-1}$. Therefore, $L_{\rm W}\sim L_{\rm
  crit}$.  Furthermore, we compare the density of the halo region
derived in equation~(\ref{eq:halo_density}) to a model calculation of
a disk-halo interaction by \citet{Norman_Ikeuchi_1989} to find that it
is within the chimney/starburst phase where the blow-out occurs.

The mass-flow rate is estimated to be $\dot{M}_{\rm flow} \sim M_{\rm
  gas}v_{\rm flow}/y \sim 8.5~\MO{\rm yr^{-1}}$, where $v_{\rm flow} =
v_s$ and $y = 15$~kpc are assumed.  For well-known bright starburst
galaxies, NGC~253 and M~82, the rates are $\dot{M}_{\rm flow} = 5.8
(v_{\rm flow}/1000~{\rm km\,s^{-1}}) (y/6.35~{\rm kpc})^{-1}~\MO{\rm
  yr^{-1}}$ \citep{Strickland_etal_2002} and $12.9 (v_{\rm
  flow}/600~{\rm km\,s^{-1}})(y/6~{\rm kpc})~\MO{\rm yr^{-1}}$
\citep{Strickland_etal_1997}, assuming the volume filling factor of
the hot plasma to be 1 and a metal abundance of $Z$=0.05~solar.  Thus,
the mass-flow rate for HCG~80a is likely to be one of the largest among
known starburst galaxies.  Furthermore, like the cylindrical
structure of the CO molecular gas observed in M~82
\citep{Nakai_etal_1987}, the outflow of cold matter undetectable in
X-rays may raise the $\dot{M}_{\rm flow}$ value significantly if it
exists.  For example, if the total mass flow is ten-times larger than
that estimated from the X-ray observation only, $t_{\rm SF}=40$~Myr
would result in a total mass loss of $\sim 4\times10^9~\MO$.  This
corresponds to about 10\% of the total galaxy mass inferred from the
$M/L_B$ ratio, $M\sim 4.7\times10^{10}~\MO$.

In conclusion it is highly plausible that HCG~80a is a starburst
galaxy that exhibits one of the most energetic outflows powered by
starburst activity known in the universe.

\subsection{Diffuse Hot Gas in HCG~80}\label{subsec:discuss_diffuse}

We obtained a severe constraint on the intensity of the diffuse
emission from the HCG~80 group region, $L_{\rm X} <
6.3\times10^{40}$~erg s$^{-1}$, which is one of the lowest among the
ROSAT groups of galaxies (figure~\ref{fig5}). The flux sensitivity of
the present {\it Chandra} observation is higher by a factor of about
25 than that of the previous ROSAT/PSPC observation
\citep{Ponman_etal_1996}.  Thus, it is clear that the current upper
limit is lower by more than one order of magnitude than that expected
from the $\sigma-L_{\rm X}$ relation.  Compared to the X-ray
luminosity of HCG~16 measured with XMM-Newton, $L_{\rm
  X}=5.0\times10^{40}h_{70}^{-2}$ erg s$^{-1}$
\citep{Belsole_etal_2003}, our upper limit is comparable to their
result.  If we further assume that the intragroup gas in HCG~80 is a
0.5~keV thermal plasma, distributed within a sphere of radius $50~{\rm
  kpc}$, the upper limits on the electron density and the total gas
mass are estimated to be $n_{\rm e} < 8.4\times10^{-4}~{\rm cm^{-3}}$
and $M_{\rm IGM}< 1.3\times10^{10}~\MO$.

If we suppose that the HCG~80 group is a virialized system, and that
the velocity dispersion properly measures the potential well, the
total mass is $M_{\rm tot}= 3\sigma_{\rm v}^2 R/G \sim 3.4\times
10^{12}~\MO$, yielding a gas mass to the total mass ratio of $<
0.004$.  This unusually small value may be a consequence of the
following possibilities: 1) HCG~80 is a chance alignment and not a
real, physical system; 2) HCG~80 is a virializing, young system and
the gas is yet to be heated to emit appreciable X-rays; or 3) the
diffuse gas is expelled from HCG~80 by, for instance, violent activity
of member galaxies.

Regarding the first possibility, the differences in the line-of-sight
velocity relative to HCG~80a are $\Delta cz= +621, +587, +145$~km
s$^{-1}$ for HCG~80b, c, d, respectively \citep{Arp_1997}.  They
satisfy the criterion $\Delta cz < 1000$~km s$^{-1}$ for the accordant
system applied in \citet{Arp_1997}.  It is admittedly difficult to
judge from the $\Delta cz$ values alone whether galaxies are indeed
concentrated compared to the field sample, since the velocity
dispersion of the member galaxies, 309 km s$^{-1}$, corresponds to a
comoving separation of 4.3 Mpc.  The four members of HCG~80, however,
are clustered within a circle of $r=50.5$~kpc on the sky, yielding a
cylindrical volume containing four galaxies of $0.07~{\rm
  Mpc^{3}}$. This is only 1/15 of the mean occupied volume, $\sim
1~{\rm Mpc^{3}}$, of field galaxies brighter than HCG~80d ($B=17$),
based on the luminosity function in the SDSS ${\rm b_j}$ band
\citep{Blanton_etal_2001}.  This is supportive of a significant galaxy
concentration in the HCG~80 group.

We further note that an exceptionally strong activity inferred in HCG
80a may be a result of galaxy interaction in the high-density
environment.  \citet{Coziol_etal_2004} quantified the level of
activity (star formation or AGN) using a sample of 91 galaxies in the
compact groups, and suggested an evolutional sequence that groups at
their early stage of evolution tend to be rich in late-type members
with high activity levels, and show a smaller velocity dispersion.
Since our results show that HCG~80 contains at least two active
galaxies, and that the velocity dispersion of $\sigma_{\rm v}=309$ km
s$^{-1}$ is relatively small among their sample, HCG~80 is likely to
be close to their configuration ``type A'', corresponding to a lower
level of evolution.  In addition, \citet{Verdes-Montenegro_etal_2001}
showed that the groups richer in early type galaxies or more compact
with larger velocity dispersion have a weak tendency to be more
deficient in H{\footnotesize I} gas than expected from the optical
luminosities, and proposed a scenario that the amount of
H{\footnotesize I} gas would decrease further with evolution by tidal
stripping and/or heating. Therefore, no significant H{\footnotesize I}
deficiency in HCG~80 may also imply the lower level of evolution. In
order to further confirm the reality of the system and to constrain
the physical properties, it is necessary to identify the distances of
the HCG~80 galaxies, and also to search for evidence of interactions.

As pointed out by \citet{Mulchaey_2000}, the hypothesis that the all
spiral-only group is a mere chance alignment is unlikely given the
existence of the our own spiral-only Local Group. We thus compare our
result with the Local Group. \citet{Wang_McCray_1993} found the soft
X-ray component with temperature 0.2 keV in the soft X-ray background,
which could be due to a warm intragroup medium in the Local
Group. \citet{Rasmussen_etal_2003} measured the absorption-line
features towards three AGNs using the XMM-Newton/RGS deep
spectroscopic data, whose redshift appear to be $z\sim0$, and placed
limits on the electron density, $n_{\rm e}<2\times10^{-4}~{\rm
  cm^{-3}}$, the scale length of the absorber, $L>140$~kpc, and its
mass, $M_{\rm IGM}< 5\times10^{10}~\MO$, in a collisional equilibrium
approximation.  Thus, the upper limit on the hot IGM in the HCG~80
group from the Chandra observation is similar to that of the Local
Group, suggesting that the spiral-only groups may contain very tenuous
IGM.

Another interpretation may be possible within the context of the
preheating models for groups (e.g., \cite{Ponman_etal_1999}). The
model predicts that the energy input through galactic winds or
outflows, powered by supernovae, should cause a more extended gas
distribution, resulting in the gas density being too low to be
detected in X-rays.  Thus, the large halo emission discovered in HCG
80a and the low density of the intergalactic medium suggested from the
analysis are not in conflict with the view of the preheating
model. However, it is inconclusive because the non-detection of
diffuse intragroup gas does not allow us to put any quantitative
constraint on a connection between the outflowing gas and the
intragroup gas. Additional follow-up observations will be meaningful
to further clarify the role of late-type galaxies in the evolution of
the IGM, probably at its early stage.

\bigskip
We are grateful to Y. Ishisaki and T. Oshima for their technical
support and useful comments.  N.O. acknowledges support from the
Special Postdoctoral Researchers Program of RIKEN.  U.M. is supported
by a Research Fellowship for Young Scientists from JSPS. 
 This research was supported in part by the Grant-in-Aid for
 Scientific Research of JSPS (14740133).

\end{document}